\documentclass[usenatbib]{mn2e}
\usepackage{natbib}  
\usepackage{epsfig}

\catcode`\@=11
\def\gta{\ifmmode{\,\mathrel{\mathpalette\@versim>\,}}
    \else{$\,\mathrel{\mathpalette\@versim>}\,$}\fi}
\def\lta{\ifmmode{\,\mathrel{\mathpalette\@versim<\,}}
    \else{$\,\mathrel{\mathpalette\@versim<}\,$}\fi}
\def\@versim#1#2{\lower 2.9truept \vbox{\baselineskip 0pt \lineskip
    0.5truept \ialign{$\m@th#1\hfil##\hfil$\crcr#2\crcr\sim\crcr}}}
\catcode`\@=12  
\def\cE{{\cal E}}
\def\Rc{R_{\rm c}}\def\vc{v_{\rm c}}
\def\Phic{\Phi_{\rm c}}
\def\Phieff{\Phi_{\rm eff}}

\def\kms{\,{\rm km}\,{\rm s}^{-1}}

\def\Myr{\,{\rm Myr}}

\def\Gyr{\,{\rm Gyr}}
\def\K{\,{\rm K}} 
\def\pc{\,{\rm pc}}
\def\kpc{\,{\rm kpc}}

\def\e{{\rm e}}
\def\d{{\rm d}}
\def\msun{\,{\rm M}_\odot}

\def\meh{\hbox{[Me/H]}}
\def\feh{\hbox{[Fe/H]}}
\def\zeh{\hbox{[$Z$/H]}}
\def\alphaH{[\alpha/\hbox{H}]}
\def\dex{\,{\rm dex}}
\def\figref#1{Fig.~\ref{#1}}
\newcommand{\beq}{\begin{equation}}
\newcommand{\eeq}{\end{equation}}

\title[Chemical evolution with radial mixing]
{Chemical evolution with radial mixing}

\author[R. Sch\"onrich and J. Binney]{Ralph Sch\"onrich$^1$\thanks{E-mail:
rasch@usm.lmu.de} and James  Binney$^2$\\
$^{1}$ Universit\"atssternwarte M\"unchen, Scheinerstr. 1, D-81679 M\"unchen, D \\
$^{2}$ Rudolf Peierls Centre for Theoretical Physics, Keble Road, Oxford OX1 3NP, UK\\}

\begin{document}

\date{6 March 2009}

\pagerange{\pageref{firstpage}--\pageref{lastpage}} \pubyear{2008}

\maketitle

\label{firstpage}

\begin{abstract}
Models of the chemical evolution of our Galaxy are extended to include radial
migration of stars and flow of gas through the disc. The models track the
production of both iron and $\alpha$ elements. A model is chosen that
provides an excellent fit to the metallicity distribution of stars in the
Geneva--Copenhagen survey (GCS) of the solar neighbourhood, and a good
fit to the local Hess diagram. The model provides a good fit to the
distribution of GCS stars in the age--metallicity plane although this plane
was not used in the fitting process.  Although this model's 
star-formation rate is monotonic declining, its disc naturally splits into an
$\alpha$-enhanced thick disc and a normal thin disc. In particular the
model's distribution of stars in the ([O/Fe],[Fe/H]) plane resembles that of
Galactic stars in displaying a ridge line for each disc. The thin-disc's
ridge line is entirely due to stellar migration and there is the
characteristic variation of stellar angular momentum along it that has been
noted by Haywood in survey data. 
Radial mixing of stellar populations with high $\sigma_z$ from inner regions of the disc to the solar neighbourhood provides a
natural explanation of why measurements yield a steeper increase of $\sigma_z$ with age than predicted by theory.
The metallicity gradient in the ISM is
predicted to be steeper than in earlier models, but appears to be in good
agreement with data for both our Galaxy and external galaxies. The models are
inconsistent with a cutoff in the star-formation rate at low gas surface
densities. The absolute
magnitude of the disc is given as a function of time in several photometric
bands, and radial colour profiles are plotted for representative times.
\end{abstract}


\section{Introduction}

Models of the chemical evolution of galaxies are key tools in the push to
understand how galaxies formed and have evolved. Their application to our
Galaxy is of particular importance both on account of the wealth of observational
data that they can be required to reproduce, and on account of the inherent 
interest in deciphering the history of our environment. 

From the pioneering papers by \cite{vdB62} and \cite{Schmidt63} it has
generally been assumed that a galaxy such as the Milky Way can be divided
into concentric cylindrical annuli, each of which evolves independently of
the others
\cite[e.g.][]{Pagel97,Chiappini97,Chiappini01,NaabO06,Colavitti08}.  The
contents of any given cylinder are initially gaseous and of extremely low or
zero metallicity. Over time stars form in the cylinder and the more massive
ones die, returning a mixture of heavy elements to the remaining gas. The
consequent increase in the metallicity of the gas and newly-formed stars is
generally moderated by an inflow of gas from intergalactic space, and, less
often, by an outflow of supernova-heated gas.

The cool, star-forming gas within any cylinder is assumed to be well mixed,
so at any time it can be characterised by a metallicity $Z(r,t)$, where $r$
is the cylinder's radius. Hence the stars formed within a given cylinder
should have metallicities $Z(r,t_{\rm f})$ that are uniquely related to their
time of formation, $t_{\rm f}$. Observations do not substantiate this
prediction; in fact \cite{Edvardsson93} showed that solar-neighbourhood stars
are widely distributed in the $(t_{\rm f},Z)$ plane -- for a detailed discussion see
\cite{Haywood06} and Section~\ref{sec:Haywood}. 

The absence of an age-metallicity relation in the solar neighbourhood is
naturally explained by radial migration of stars
\citep{SellwoodB,Haywood08,Roskar2}.
It has been recognised for many years that scattering by spiral structure and
molecular clouds gradually heats the stellar disc, moving stars onto ever
more eccentric and inclined orbits. Stars that are on eccentric orbits
clearly contribute to different cylindrical annuli at different phases of
their orbits, and thus tend to modify any radial gradient in the metallicities
of newly formed stars.  Moreover, scattering events also change the guiding
centres of stellar orbits, so even a star on a circular orbit can be found at
a different radius from that of its birth. In fact, \cite{SellwoodB} argued
that the dominant effect of transient spiral structure is resonant scattering
of stars across the structure's corotation resonance, so even a star that is
still on a near-circular orbit may be far from its radius of birth.
\cite{Roskar1} showed that in a cosmological
simulation of galaxy formation that included both stars and gas, resonant
scattering at corotation caused stars to move outwards and gas inwards, with
the result that the stellar disc extended beyond the outer limit of star
formation; the outer disc was entirely populated by stars that had formed
much further in and yet were still on nearly circular orbits. This simulation
confirmed the conjecture of \cite{SellwoodB} that gas would participate in
resonant scattering alongside stars.  

We distinguish two drivers of radial migration: when the angular momentum of
a star is changed, whether by scattering at an orbital resonance or by
non-resonant scattering by a molecular cloud, the star's guiding-centre
radius changes and the star's entire orbit moves inwards or outwards
depending on whether angular momentum is lost or gained. When a scattering
event increases a star's  epicycle amplitude without changing its angular
momentum, the star contributes to the density over a wider range of radii. In
a slight modification of the terminology introduced by \cite{SellwoodB}, we
say that changes in angular momentum cause ``churning'' while changes
in epicycle amplitude lead to ``blurring''. This paper extends models of
Galactic chemical evolution to include the effects of churning and blurring.

Given the strength of the arguments that cold gas should participate in
churning alongside stars, and that shocks induced by spiral structure cause
gas to drift inwards, it is mandatory simultaneously to extend traditional
chemical evolution models to include radial flows of gas within the disc.
\cite{LaceyF85} studied chemical evolution in the presence of a radial inflow
of gas and demonstrated that a radial flow enhances the metallicity gradient
within the disc. This enhancement plays an important role in our models,
which differ from those of Lacey \& Fall in that they include both radial gas
flows and radial migration of stars. Moreover we can fit our models to
observational data that is much richer than that available to \cite{LaceyF85}.

Our models are complementary to ab-initio models of galaxy formation such as
those presented by \cite{SamlandG} and \cite{Roskar2} in that they allow the
solar neighbourhood to be resolved in greater detail, and because they are
enormously less costly numerically, they permit parameter searches to be made
that are not feasible with ab-initio models.

The paper is organized as follows. Section \ref{sec:goveqs} presents the
equations upon which the models are based. These consist of the rules that
determine the rate of infall of fresh gas, the rate of star formation,
details of the stellar evolution tracks and chemical yields that we have used
and descriptions of how churning and blurring are implemented. Section
\ref{sec:standard} describes in some detail a ``standard'' model of the
evolution of the Galactic disc. This covers its global properties but focuses
on what would be seen in a survey of the solar neighbourhood. Section
\ref{sec:Snhd} presents the  details of the selection
function that is required to mimic the Geneva--Copenhagen sample (GCS) of
solar-neighbourhood stars published by \cite{Nordstrom04} and
\cite{HolmbergNA}, and explains how
this sample has been used to constrain the model's parameters. Section
\ref{sec:trend} explains how the observable properties of the model depend on
its parameters.  Section \ref{sec:reln} discusses the relation of the present
models to earlier ones, and discusses the extent to which it is consistent
with the analysis of solar-neighbourhood data by \cite{Haywood08}. Section
\ref{sec:conclude} sums up.

\section{Governing equations}\label{sec:goveqs}

The simulation is advanced by a series of discrete timesteps of duration
$30\Myr$. 

The disc is divided into 80 annuli of width $0.25\kpc$ and central radii that
range from $0.125\kpc$ to $19.875\kpc$.  In each annulus there is both
``cold'' ($\sim30\K$) and ``warm'' ($\gta10^4\K$) gas with specified
abundances $(Y,Z)$ of helium and heavy elements. The ``warm gas'' is not
available for star formation and should be understood to include both
inter-cloud gas within the plane and extraplanar gas, which probably contains
a significant fraction of the Galaxy's ISM. Indeed, in NGC$\,$891, a galaxy
similar to the Milky Way, of order a third of HI is extraplanar
\citep{OosterlooFS}. In the Milky Way this gas would constitute the
``intermediate-velocity clouds'' that are observed at high and intermediate
Galactic latitudes \citep{Kalberla08}. 

Within the heavy elements we keep track of the abundances
of O, C, Mg, Si, Ca and Fe. Each annulus has a stellar population for each
elapsed timestep, and this population inherits the abundances $Y$, $Z$, etc.,
of the local cold gas.  At each stellar mass, the stellar lifetime is
determined by the initial abundances, and at each age we know the luminosity
and colours of such of its stars that are not yet dead.  Each stellar
population is at all times associated with the annulus of its birth; the
migration of stars is taken into account as described below only when
returning matter to the ISM or constructing an observational sample of stars.

\subsection{Metallicity scale}\label{subsec:metscale}

The whole field of chemical modelling has been thrown into turmoil by the
discovery that three-dimensional, non-equilibrium models of the solar
atmosphere require the metal abundance of the Sun to be
$Z_\odot=0.012-0.014$ \citep{Grevesse07} rather than the traditional value
$\sim0.019$. This work suggests that the entire metallicity scale needs to be
thoroughly reviewed: if the Sun's metallicity has to be revised downwards,
then so will the metallicities of most nearby stars. Crucially there is the
possibility that values for the metallicity of the ISM require revision: some
values derive from measurements of the metallicities of short-lived stars
such as B stars and require downward revision \citep[e.g.][]{Cunha}, while
others are inferred from measurements of the strengths of interstellar
emission lines, and are not evidently affected by changes in stellar
metallicities.  If the metallicity scale of stars were lowered while that of
the ISM remained substantially unaltered, it would be exceedingly hard to
construct a viable model of the chemical evolution of the solar
neighbourhood. Moreover, both the stellar catalogue and most of the
measurements of interstellar abundances with which we wish to compare our models
are on the old metallicity scale, and unphysical anomalies will become rife
as soon as one mixes values on the old scale with ones on the new. Therefore
for consistency we use the old solar abundance $Z_\odot=0.019$ and exclude
from considerable metallicity values that are on the new scale.

\subsection{Star-formation law}

Stars form according to the \cite{Kennicutt98} law. Specifically, with the
surface density of cold gas $\Sigma_{\rm g}$ measured in $\msun\pc^{-2}$ and
$t$ in Myr, star formation increases the stellar surface density at a rate
 \begin{equation}\label{eq:Kenni}
{\d \Sigma_*\over\d t}= 1.2\times10^{-4}\cases{
\Sigma_{\rm g}^{1.4}&for $\Sigma_{\rm g}>\Sigma_{\rm crit}$\cr
C\Sigma_{\rm g}^4&otherwise,}
\end{equation}
 where the threshold for star formation, $\Sigma_{\rm crit}$ is a parameter
of the model and $C = \Sigma_{\rm crit}^{-2.6}$ ensures that
the star-formation rate is a continuous function of surface density. The
normalisation in equation (\ref{eq:Kenni}) was chosen to yield the observed
surface densities of gas and stars near the Sun.

The stars are assumed to be distributed in initial mass over the range
$(0.1,100)\msun$ according to the Salpeter function, $\d N/\d M\propto
M^{-2.35}$. The luminosities, effective temperatures, colours and lifetimes
of these stars are taken by linear interpolation in $(Y,Z)$ from the values
given in the BASTI database \citep{Cassisi06}.

\subsection{Return of metals}

The nucleosynthetic yields of individual metals are in many cases still
subject to significant uncertainties \cite[e.g.][]{Thomas98}; in fact models
of the chemical evolution of the solar neighbourhood have been used to
constrain these yields \citep{Francois04}.

For initial masses in the ranges $5-11\msun$ and $35-100\msun$ values of
$X,Y,Z$,C and O were taken from \cite{Maeder92} using a non-linear
interpolation scheme: the paper gives yields $Y_{\rm LZ}$ for a low
metallicity ($Z=10^{-4}$)
and yields $Y_{\rm HZ}$ for a high metallicity ($Z=0.02$).
Guided by the metallicity-dependence of the
sizes of CO cores reported by \cite{Portinari98} we take
 \begin{equation}
Y(Z)=(1-\alpha)Y_{\rm
LZ}+\alpha Y_{\rm HZ},
\end{equation} where
 \begin{equation}
\alpha=\cases{0&for $Z<0.005$\cr
320(Z-0.005)&for $0.005<Z<0.0075$\cr
0.8+16(Z-0.0075)&for $0.0075<Z<0.02$\cr
1& for $Z>0.02$}
\end{equation}
 The yields of elements other than $X,Y,Z$,C and O from stars with masses in
this range were taken from the {\it ORFEO} database of \cite{LimongiC} with
the mass cut set such that $0.05 \msun$ of $^{56}$Ni is produced; this
relatively low mass cut reproduces the Ca/Fe ratio measured in very
metal-poor stars by \cite{Lai08}.  Stars less massive than $10\msun$ were
assumed to produce no elements heavier than O. For stars with masses
$<5\msun$, the yields were taken by linear interpolation from
\cite{Marigo01}. 

For initial stellar masses in the range $11-35\msun$ we used the
metallicity-dependent yields of heavy elements from \cite{ChieffiL} by linear
interpolation on mass and metallicity, extrapolating up to $\alpha=1.5$ or
$Z=0.03$ respectively.  \cite{ChieffiL} used a relatively
high mass cut, which produced $0.1 \msun$ of $^{56}$Ni. With our
interpolation  the average
amount of $^{56}$Ni produced is well within the expected range.

A fraction $f_{\rm eject}$ of the gas ejected by dying stars leaves the
Galaxy; we tested models with $0\le f_{\rm eject}\le0.05$ \citep{Pagel97}.
Increasing $f_{\rm eject}$ has the effect of reducing the final metallicity
of the disc; in fact there is almost complete degeneracy between the values
of $f_{\rm eject}$ and nucleosynthetic yields. In view of the evidence that
star formation near the Galactic centre drives a Galactic wind
\citep{BlandHCohen}, we set $f_{\rm eject}=0.15$ at $R<3.5\kpc$ in models that
use the accretion law (\ref{eq:Mdot}) below. At all other radii we set
$f_{\rm eject}=0.04$.

A fraction $f_{\rm direct}$ of the ejecta goes straight to the cold gas
reservoir of the local annulus, and the balance goes to the annulus's
warm-gas reservoir. Setting $f_{\rm direct}$ to values $\sim0.2$ has a
significant impact on the number of extremely metal-poor stars predicted near
the Sun. However, such large values of $f_{\rm direct}$ are not well
motivated physically, and in the models presented here $f_{\rm direct}=0.01$
has a negligible value. 

In each timestep $\delta t$ a fraction $\delta t/t_{\rm cool}$ of the
``warm'' gas (which includes extraplanar gas) transfers to the cold-gas
reservoir from which stars form.  The parameter $t_{\rm cool}$ is determined
by the dynamics of extraplanar gas and the balance between radiative cooling
and shock heating within the plane. Consequently, its value cannot be
determined a priori from atomic physics.  Increasing $t_{\rm cool}$ increases
the mass of ``warm'' gas and delays the incorporation of freshly made metals
into new stars, so the number of very metal-poor stars formed
increases with $t_{\rm cool}$.  Although of order a quarter of the neutral
hydrogen of NGC 891 is extraplanar \citep{OosterlooFS}, some of this gas will
be formerly cold interstellar gas that has been shock accelerated by stellar
ejecta. We do not model shock heating of cold gas, and replenish the warm-gas
reservoir exclusively with stellar ejecta (from stars of every mass). Hence
the warm-gas reservoir should be less massive than the sum of the extraplanar
and warm in-plane bodies of gas in a galaxy like NGC 891.  We have worked
with values $t_{\rm cool}\gta1\Gyr$ that yield warm-gas fractions of order 10
percent.  The results of the models are not sensitive to the value of $t_{\rm
cool}$.

There is abundant evidence that pristine intergalactic gas disappeared from
the intergalactic medium (IGM) long ago: quasar absorption-line studies
reveal an early build up of heavy elements in the IGM \citep{Pettini}.  While
it is clear that the disc formed from material that had been enriched by
pregalactic and halo stars, it is unclear what abundances this material had.
We take the chemical composition of the pre-enriched gas to be that of the
``warm'' ISM after two timesteps, starting with $5\times10^8\msun$ of
pristine gas.  In each of the following four timesteps, a further
$1.25\times10^8\msun$ of gas with this metallicity is added to the disc. The
surface density of the added gas is proportional to
 \begin{equation}
(1.0-\e^{(R-19.8\kpc)/11.8\kpc})\e^{-R/4\kpc}.
\end{equation}
 Thus the surface density is exponential with scalelength $4\kpc$ inside
$\sim R_0$ but tapered to zero at the outer edge of the grid. The existence
at the outset of a warm, pre-enriched component of the ISM is physically well
motivated and proves the most effective way of producing the right number of
metal-poor stars.

\begin{figure}
\epsfig{file=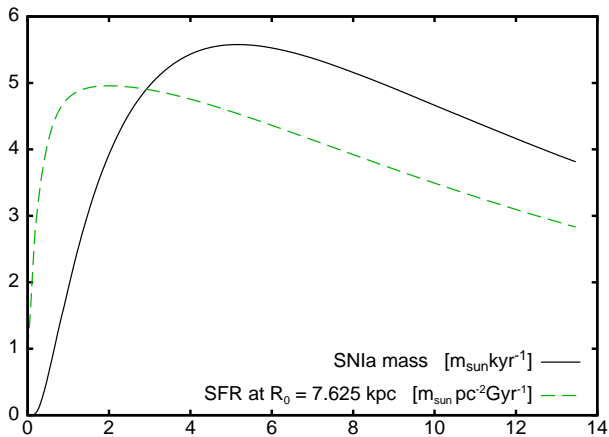,angle=-90,width=\hsize}
\caption{The rate of mass injection by SNIa in the standard model (solid black
  line) versus time. The broken green line gives the star formation rate in the solar annulus.
\label{fig:SNIrate}}
\end{figure}

Type Ia supernovae are included by assuming that $7.5$ per cent of the mass in white
dwarfs formed by stars of initial mass 3.2 to $8.5\,\msun$ ultimately
explodes in type Ia supernovae. The yields were taken to be those of the W70
model in \cite{Iwamoto99}. It is believed that the progenitors of type Ia
supernovae have lifetimes of order a Gyr \citep{Forster06}  and we have
taken the mass $M_{\rm WD}$ of the population that survives to time $t$ from
white-dwarf formation to satisfy
 \begin{equation}\label{eq:SN}
{\d M_{\rm WD}\over\d t}=\cases{0&for $0<t\le0.15\Gyr$\cr
-M_{\rm WD}/1.5\Gyr&for $T>0.15\Gyr$.}
 \end{equation} 
 The rate of type Ia SNe is constrained by the requirements that (i) [O/Fe]
has to fall from $\sim0.6$ for the oldest stars to around $-0.1$, and (ii)
[Ca/Fe] should go from about $\sim0.3$ to $\sim0$.  The full curve in
Fig.~\ref{fig:SNIrate} shows for the best-fitting model the
mass-return rate as a function of time, while the green dashed curve shows the SFR
at the solar radius.

\subsection{Inflow}

It is generally agreed that viable models of galactic chemical evolution
require the disc to be constantly fed with gas from intergalactic space;
inflow resolves several serious problems, including (i) the appearance of too
many low-metallicity stars near the Sun \citep[the ``G-dwarf problem'',
e.g.][]{Pagel97}, (ii) excessive metallicity of the current ISM, (iii) an
unrealistically low abundance of deuterium in the current ISM
\citep{Linsky06}. Moreover, both
the short timescale for the current ISM to be consumed by star formation and
direct manifestations of infalling gas \citep{Sancisi08} argue strongly for
the existence of infall. Unfortunately, many aspects of infall are extremely
uncertain. We find that the predictions of our models depend
sensitively on how these uncertainties are resolved, so to the extent that
other aspects of our models have sound foundations, they can usefully
constrain the nature of infall. 

In principle the rate and radial distribution of infall is determined by
cosmology. For example, \cite{NaabO06} infer it by assuming that the disc
scale length grows in parallel with the cosmic scale, while
\cite{Colavitti08} derive the global rate from N-body simulations. At this
stage we feel that cosmological simulations are beset by too many
uncertainties to deliver even a secure global infall rate, never mind the
radial distribution of infall. In particular the extent of angular-momentum
exchange between baryons and dark matter is controversial, as are the extent
to which gas is accreted from cold infall rather than a hot corona. Moreover,
nothing is known with any confidence about the dynamics of the corona.

For want of clear inputs from cosmology we have sought a flexible
parametrisation of infall. First we parametrise the global infall rate, and
then the radial distribution of infall.

\subsubsection{Infall rate}

We have investigated two approaches to the determination of the infall rate.
The first starts with a quantity of gas
($8\times10^9\msun$) and feeds gas into it
at a rate
 \begin{equation}\label{eq:Mdot}
\dot M={M_1\over b_1}\e^{-t/b_1}+{M_2\over b_2}\e^{-t/b_2}.
\end{equation}
 Here $b_1\simeq0.3\Gyr$ is a short timescale that ensures that the
star-formation rate peaks early on, while $b_2\simeq14\Gyr$ is a long
timescale associated with sustained star formation in the thin disc. We adopt
$M_1\simeq4.5\times10^9\msun$ and choose $M_2$ such that after $12\Gyr$ the
second exponential has delivered $2.6\times10^{10}\msun$.

In an alternative scheme, the gas mass within the disc is determined a priori
and infall is assumed to be available to maintain the gas mass at its
prescribed level. We have investigated schemes in which the gas mass declines
exponentially with time, but focused on models in which it is held
constant at $8.4\times10^9\msun$; models in which the gas mass declines
exponentially produce very similar results to models in which the infall rate
declines exponentially.

\subsubsection{Distribution of infall}\label{sec:infall}

We know even less about the radial distribution of the infalling gas than we
do about the global infall rate. In fact our only constraint is that the
stellar disc
has an approximately exponential surface density now, and was probably
exponential at earlier times too. Besides the
star-formation law, the structure of the stellar disc depends on both the
radial distribution of infall and gas flows within the disc, and a disc that
is consistent with observations will not be formed if either the radial
infall profile or the internal gas flow is fixed without regard to the other
process.  Consequently, the requirement that only observationally acceptable
discs be produced requires one to develop a parametrisation that couples
infall and flow in a possibly unphysical way. The scheme we have developed
involves such an unphysical coupling -- this is the price one pays for a
scheme that allows one to explore as economically but fully as possible a
range of infall profiles and internal flows that are consistent with the
known radial structure of the disc.

We start from the assumption that the surface-density of gas is at all times
exponential, $\Sigma_{\rm g}(R)\propto\e^{-R/R_\d}$, where $R_\d=3.5\kpc$ is
chosen such that with the star-formation law adopted above, the inner stellar
disc acquires a scale length $R_*=R_\d/1.4=2.5\kpc$ similar to that
determined from star counts \citep{Robin03,Juric08}. Our value for the scale
length of the gas disc is in good agreement with the value measured by
\cite{Kalberla08}: $3.75\kpc$.  Notice that we assume not only that the
stellar disc is exponential, but that its scale length is unchanging. Hence
we are assuming that the disc forms simultaneously at all radii, rather than
``inside-out''. The remarkably large age estimated for the solar
neighbourhood (at $R=3R_*$) suggests simultaneous formation \citep[][ and
references therein]{AumerB08}. However, our work could be readily generalised
to inside-out growth by making $R_\d$ a specified function of time
\citep[e.g.][]{NaabO06}, but we reserve this extension for a later paper.

Our scheme for parameterising infall and flow depends on two parameters,
$f_{\rm A}$ and $f_{\rm B}$ and is easiest to explain by considering first
the limiting cases in which one parameter vanishes.

Either of the algorithms of the last subsection specifies what the total gas
mass should be at the start of a timestep: this is either the prescribed
constant or, when equation (\ref{eq:Mdot}) is used, it is the mass in the
disc at the end of the previous timestep plus the amount that falls in during
the most recent timestep. Hence the mass that should be in each annulus at
the start of a timestep follows from the assumed exponential profile of the
gas disc.  Subtracting from this the mass that was present after the previous
timestep, we calculate the need, i.e. the amount of gas that has to be added,
of the $i$th annulus $\Delta M_i$. 

\begin{figure}
\epsfig{file=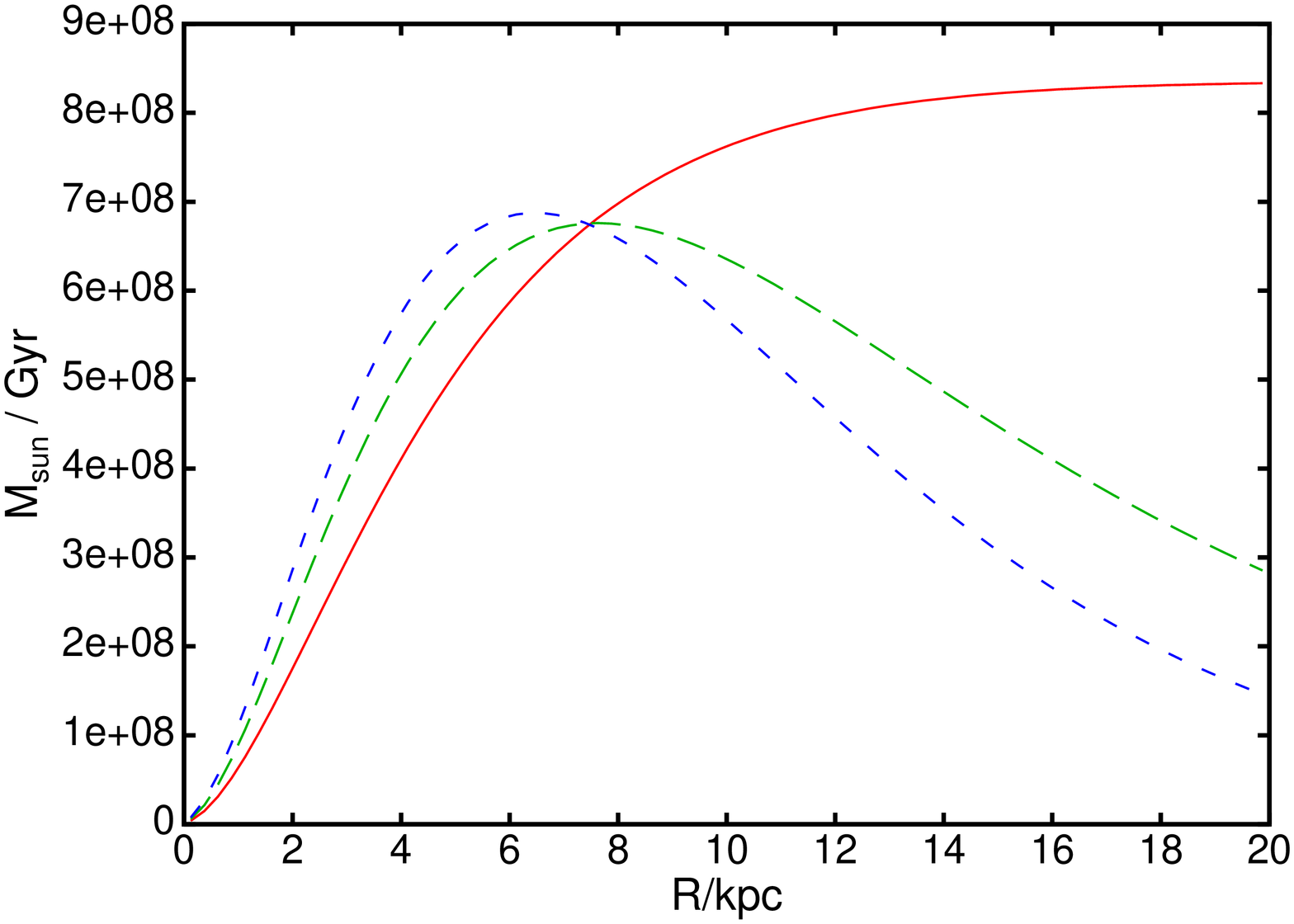,angle=-90,width=\hsize}
\epsfig{file=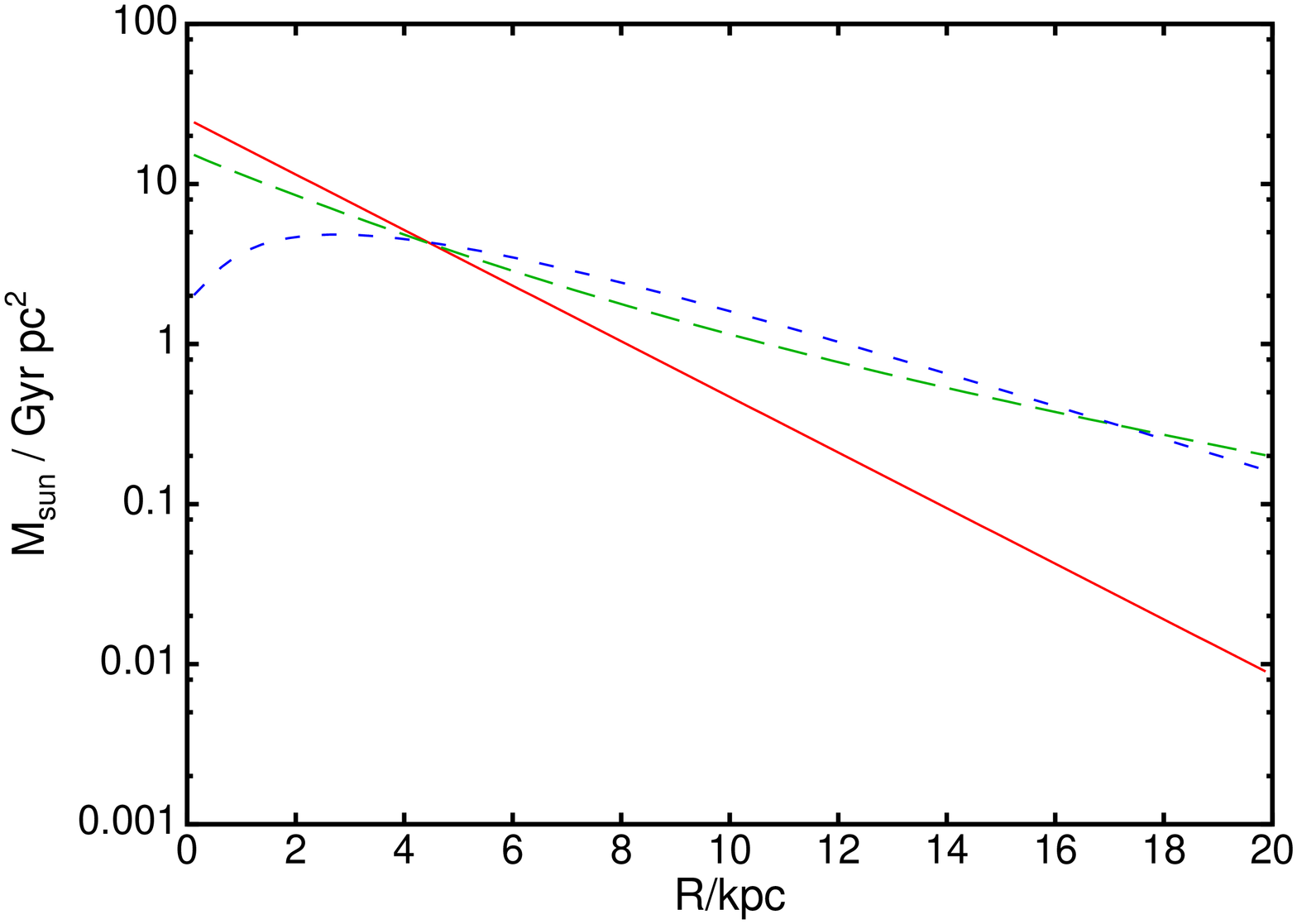,angle=-90,width=\hsize}
 \caption{Upper panel:
The rate of
flow of gas over the circle of radius $R$ induced by  infall Scheme A with $f_{\rm
A}=0.6$ (red curve), infall Scheme B with $f_{\rm B}=0.05$ (blue short dashed
curve), and infall Scheme AB with $f_{\rm A}=0.35$ and $f_{\rm B}=0.025$ (green
long dashed curve; the standard model).   Lower panel: the corresponding
rates of accretion from the IGM per unit area of the disc.}
 \label{fig:drag}
\end{figure}

We fill annuli up with gas in sequence, starting with the innermost ring $0$.
When $f_{\rm B}=0$ (``Scheme A'') this annulus receives $f_{\rm A}\Delta M_0$
from the IGM, and grabs the balance, $(1-f_{\rm A})\Delta M_0$, from annulus
1, where $f_{\rm A}\simeq0.2$ is a parameter of the model. Annulus $1$
receives $f_{\rm A}\Delta M_1$ from the IGM, and grabs the balance of its
requirement, $(1-f_{\rm A})(\Delta M_1+\Delta M_0)$, from annulus $2$. The
updating of every annulus proceeds similarly, until the last annulus is
reached, which covers its entire need from the IGM. The characteristic of
this ``Scheme A'' is the development of a large flux of gas through the outer
rings -- an example is given by the full red curve in the upper panel of
\figref{fig:drag}.  This flux transports inwards metals synthesised in these
rings and tends to deposit them at intermediate radii, where the inward flux
is diminishing.

When $f_{\rm A}=0$ (``Scheme B'') annulus $0$ obtains $f_{\rm B}\Delta M_0$
from the IGM and the rest from annulus $1$. Annulus $1$ now obtains $f_{\rm
B}[\Delta M_1+(1-f_{\rm B})\Delta M_0]$ from the IGM, and so on to the
outermost ring, which is again entirely fed by the IGM. The short dashed blue
curve in the upper panel of \figref{fig:drag} shows a typical example of a
mass flow through the disc with Scheme B. Whereas the flow generated by
Scheme A (red curve) increases monotonically from the centre, the Scheme-B
flow rises quickly with galactocentric distance $R$ near the centre but then
peaks at $R\simeq 5\kpc$. In the outer region in which the inflow is small,
the metallicity forms a plateau.  The extent of this plateau is controlled by
$f_{\rm B}$: the larger $f_{\rm B}$, the smaller the radius at which the
inflow rate peaks and the further in the metallicity plateau extends.

In either of these schemes a fixed fraction of each annulus's need is taken
from the IGM, but the definition of ``need'' is different in the two schemes:
in Scheme B it includes the gas that was taken from it by its inner
neighbour, and in Scheme A it does not. In Scheme A only a fixed fraction of
the local need is provided by the IGM, so the flow $F_r$ in the disc
continuously builds up through the disc. 
In Scheme B, by contrast, a part of
the flow required in Scheme A is met by additional accretion. Consequently, if one wrote
an equation for $\d F_r/\d r$, a term  $-(f_B\Delta r) F_r$ would appear,
where $\Delta r = 0.25\kpc$ is the width of annuli, and this term
drives exponential decay of $F_r$.
Scheme A enhances the metallicity of the middle section of the disc
and causes the metallicity gradient to be steepest towards the outside of the
disc, Scheme B enhances the metallicity of the inner disc and flattens the
gradient at large radii.

In Scheme A, if $f_{\rm A}$ is set too low, the flux of gas through the outer
annuli becomes implausibly large in relation to the mass of gas that is in
these annuli, and radial flow velocities $v_R\gta20\kms$ are predicted. In
Scheme B $f_{\rm B}$ can be quite small because, although the flow of gas
through the disc builds up more quickly at small radii, it peaks at a few
kiloparsecs and then declines to small values in the outer disc.  If either
$f_{\rm A}$ or $f_{\rm B}$ is large, the flow through the disc becomes small
and the metallicity of the solar neighbourhood becomes unrealistically large
through the accumulation of metals created at the solar radius and beyond.

 Satisfactory fits to the data can be obtained only when both $f_{\rm A}$ and
$f_{\rm B}$ are non-zero In this ``Scheme AB'' annulus $0$ receives a mass
$(f_{\rm A} + f_{\rm B})\Delta M_0$ from the IGM and grabs the balance
$M_{01} = (1-f_{\rm A} - f_{\rm B}) \Delta M_0$ from annulus $1$. Annulus $1$
receives a mass $f_{\rm A} \Delta M_1 + f_{\rm B}(\Delta M_1 + M_{01}) $ from
the IGM and grabs the balance of its requirement from annulus $2$, and so on.
Notice that the radial flow profile in Scheme AB is not simply the sum of the
corresponding profiles for Schemes A and B used alone. The green curve in the
upper panel of \figref{fig:drag} shows the radial flow profile obtained with
Scheme AB with the parameters of the standard model. In this model the radial
velocity of disc gas currently rises roughly linearly from zero at the centre to
$1.3\kms$ at the Sun. Beyond the Sun a plot of radial velocity versus radius
gradually steepens to reach $5\kms$ at the edge of the disc.

For each accretion scheme, the lower panel of \figref{fig:drag} shows the
corresponding radial distribution of accretion from the IGM.

\subsubsection{Metallicity of the IGM}

We have to prescribe the metallicity and alpha-enhancement of gas taken from
the IGM. It is far from clear how this should be done.

Quasar absorption line-studies reveal an early build up of heavy elements in
the IGM \citep{Pettini}. Moreover, the handful of high-velocity clouds for
which metallicities have been measured, have heavy-element abundances of
order a tenth solar \citep{Wakker}. Finally, the metallicities of the most
metal-poor thick-disc stars are similar to the metallicities of the most
metal-rich halo stars, which suggests that the early disc was pre-enriched by
pregalactic and halo stars. We assume that throughout the simulation accreted
gas has metallicity $Z=0.1Z_\odot$.

Given that the thick disc is alpha-enhanced \citep{Venn04}, it is clear that
when disc formation starts, infalling gas must be alpha-enhanced. It is
natural that this enhancement should decline with time as Fe from  type Ia
SNe finds its way into the IGM. Indeed, in addition to gas that flows out in
the Galactic wind \citep{BlandHCohen}, type Ia SNe in dwarf spheroidal
galaxies will have contributed their Fe to the local IGM, and if the
Magellanic Stream is made of gas torn from the SMC, it will have been
enriched with Fe from SNe in the SMC. Thus we expect the metallicity and
alpha enhancement of the IGM to be time dependent and governed by the
chemical-evolution histories of galaxies.

These considerations suggest making the $\alpha$-enhancement of the IGM
reflect that of an outer annulus of the Galaxy; the chemical evolution of
this ring acts as a proxy for the combined chemical evolution of the many
contributors to the chemical evolution of the IGM. If the IGM were assumed to
mirror the outermost ring, its $\alpha$-enhancement would remain extremely
low because this annulus takes all its gas from the IGM, and passes what few
heavy elements it synthesises inwards.  Hence the IGM must mirror an outer
annulus but not the outermost. In our models the $\alpha$-enhancement of the
IGM mirrors the annulus with radius $R=12.125\kpc$.  Since yields of $\alpha$
elements decline with increasing metallicity, the outer disc should be
$\alpha$-enhanced.

\subsection{Churning}

Transient spiral arms cause both stars and gas to be exchanged between annuli
in the vicinity of the corotation resonance.  Such exchanges automatically
conserve both angular momentum and mass. Since these exchanges are driven by
spiral structure, in which hot and extraplanar gas is not expected to
participate, churning is confined to stars and cold gas.  We restrict
exchanges to adjacent rings but allow two exchanges per timestep, so within a
timestep second-nearest neighbouring rings exchange mass. 

Further studies of spiral structure in high-quality N-body simulations are
required to determine how the probability of a star migrating varies across
the disc. In the absence of such studies the following dimensional argument
suggests what the answer might be. Consider the probability $P_{\rm ex}$ that
in a characteristic dynamical time $\kappa^{-1}$ (where $\kappa$ is the local
epicycle frequency) a star is involved in a resonant exchange across
corotation. It is natural that a process dependent on gravitational
self-energy in the disc should scale with the square of the surface density.
Toomre's $Q=\sigma\kappa/\pi G\Sigma$, where $\sigma$ is the radial velocity
dispersion, is a dimensionless variable, so we conjecture that $P_{\rm
ex}\propto1/Q^2$. Our grid is uniform in $R$ whereas an exchange across
corotation changes $R$ by of order the most unstable wavelength $\lambda_{\rm
crit}=\sigma Q/\kappa$. The number of swaps between rings required to wander
a distance $\lambda_{\rm crit}$ scales as $\lambda_{\rm crit}^2$. Moreover, the number of
ring-swaps in time $\kappa^{-1}$ scales as $\kappa^{-1}$, so the ring-swap
probability per timestep $P_{\rm ring}$ should be $\lambda_{\rm crit}^2\kappa$ times
$P_{\rm ex}$. This argument yields $P_{\rm ring}\propto\sigma^2/\kappa$. In
realistic cases $\sigma^2\propto\Sigma$ and $\kappa\propto R$, so $P_{\rm
ring}\propto \Sigma R\propto M$, the mass of a ring. This argument suggests
that we take the probability $p_{ij}$ that in a given half-timestep a star or
gas cloud in the $i$th annulus is transferred to the $j$th annulus to be
 \begin{equation}\label{eq:churn}
p_{ij}=\cases{k_{\rm ch} M_j/M_{\rm max}&for $j=i\pm1$\cr0&otherwise,}
\end{equation}
 where $M_j$ is the mass in cold gas and stars in the $j$th annulus and
$M_{\rm max}=\max_j(M_j)$.  This rule ensures that the mass transferring
outwards from the $i$th annulus is proportional to $M_iM_{i+1}$, and an equal
mass transfers inwards, ensuring that angular momentum is conserved. The
constant $k_{\rm ch}$ is the largest transition probability for any annulus
in a given timestep. It is treated as a free parameter to be fitted to the
data.  

\begin{figure}
\epsfig{file=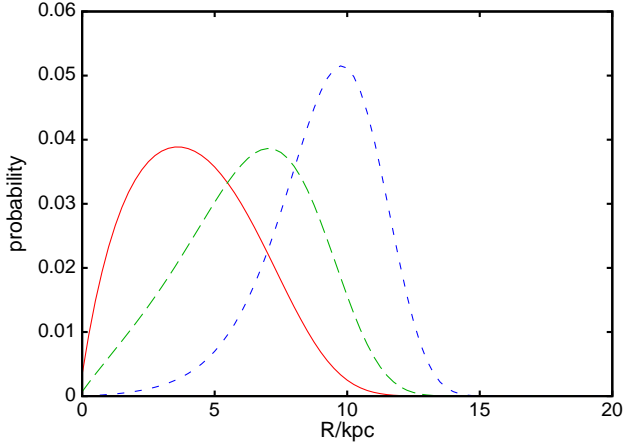,angle=-90,width=\hsize}
\caption{The radial distribution of the guiding centres of  $12\Gyr$-old
stars that were born at 5 (red), 7.6 (green, long dashed)
and $10\kpc$ (blue, short dashed) when the churning fraction $k_{\rm
ch}=0.25$.} 
\label{fig:churndist}
\end{figure}

The procedure for distributing the
metals released by a population of stars born in annulus $i$ is as follows.
The probability that a star born in annulus $i$ at timestep $m$ is found to
be in annulus $j$ at timestep $n$ is equal to the $ij$th element of the
product matrix ${\bf p}_{m}{\bf p}_{m+1}\times\cdots{\bf p}_{n}$. In practice
we recompute ${\bf p}$ only each five timesteps and approximate ${\bf
p}_{m}\times\cdots{\bf p}_{m+4}$ by ${\bf p}_{m}^5$. Fig.~\ref{fig:churndist}
shows the extent to which the guiding centres of stars are changed over the
lifetime of the Galaxy when $k_{\rm ch}=0.25$. 

\subsection{Blurring}

In addition to changing their guiding-centre radii through the churning
process, stars oscillate around their guiding centres with steadily
increasing amplitudes. Consequently, stars spend time away from their
guiding-centre radii. For simplicity, we assume in this section that the
circular speed $\vc$ is independent of radius and that the  vertical motion
can be ignored because it decouples from motion in the plane.

The fraction of its time that the orbit with energy and
angular momentum $E,L$ spends in a radial interval $(R,R+\d R)$ is
 \begin{equation}
\d p={\d t\over T}={1\over T}{\d R\over v_R}=
{\Omega_R\over\pi}{\d R\over\sqrt{2(E-\Phieff)}},
\end{equation}
 where $T\equiv\pi/\Omega_R$ is the half period and $\Phi_{\rm
eff}(R,L)\equiv\Phi(R)+L^2/2R^2$ is the effective potential. We need to
average this over all stars with given $L$. These stars have some
distribution over the energy $E=\frac12 v_R^2+\Phieff$. It is expedient to
decompose $E$ into the energy $\Phic(L)\equiv\Phi(\Rc)+L^2/2\Rc^2$ of the
circular orbit (with radius $R_c$) of angular momentum $L$ and the random energy $\cE\equiv
E-\Phi_{\rm c}$. Following \cite{ShuDF} we take the distribution function (DF) to
be 
\begin{equation}\label{eq:DF}
f(\cE,L)={F(L)\over\sigma^2}\e^{-\cE/\sigma^2},
\end{equation}
 where $F(L)$ is a function to be determined. The DF (\ref{eq:DF}) ensures
that the radial velocity dispersion is approximately (but not exactly)
$\sigma$. Normalizing $f$ such that $\int\d L\,\d J_R \,f=1$, where
$J_R(\cE)$ is the radial action, the probability that a randomly chosen star
lies in $(R,R+\d R)$ is $\int\d L\,\d J_R\,f\d p$. Recalling that $\d L\,\d
J_R=\d L\,\d\cE/\Omega_R$ and substituting for $f$ and $\d p$, we find that
the number of stars in the annulus is
 \begin{eqnarray}
\d n(R)&=&\int\d L\,\d J_R\,(f\d p)\nonumber\\
&=&{N\d R\over\pi}\int\d L\,{F\over\sigma^2}\int_{\Phieff-\Phic}^\infty
\d\cE\,{\e^{-\cE/\sigma^2}\over\sqrt{2(\cE+\Phic-\Phieff)}}\\
&=&
{N\d R\over\sqrt{2}\pi}\int\d L\,{F\over\sigma^2}\e^{[\Phic-\Phieff]/\sigma^2}\int_0^\infty
\d x\,
{\e^{-x/\sigma^2}\over\sqrt{x}},\nonumber
\end{eqnarray}
 where $N$ is the total number of stars in the system. The integral over $x$
is simply $\sigma\int\d t\,\e^{-t}/\surd t=\sqrt{\pi}\sigma$.
Thus we can conclude that the probability per unit area associated with a
star of given $L$ is
 \begin{equation}\label{givesP}
P(R)={\d n\over N2\pi R\d R}={K\over\sigma
R}\,\exp\left[{\Phic(L)-\Phieff(R,L)\over\sigma^2}\right],
\end{equation}
 where $K$ is chosen such that $1=2\pi\int\d R\,RP(R)$.

The parameter $\sigma$ used in these formulae is actually smaller than the
rms radial velocity dispersion, which is given by
 \begin{equation}\label{givesVR}
\langle v_R^2\rangle={\sqrt{2\pi}\over R\Sigma}
\int\d L\, F
\sigma\exp[(\Phi_{\rm c}-\Phi_{\rm eff})/\sigma^2],
\end{equation}
 where the stellar surface density is
 \begin{equation}\label{givesSigma}
\Sigma(R)= {\sqrt{2\pi}\over R}\int\d L\, {F\over\sigma}
\exp[(\Phi_{\rm c}-\Phi_{\rm eff})/\sigma^2]. 
\end{equation}
 For specified radial dependencies of $\Sigma$ and $\langle v_R^2\rangle$,
equations (\ref{givesVR}) and (\ref{givesSigma}) can be used to determine the
functions $F(L)$ and $\sigma(L)$ \citep{DehnenDF}. However, in the present
application it is not $\Sigma(R)$ that we wish to specify, but the number of
stars with guiding centres in each ring:
 \begin{eqnarray}\label{givesNL}
{\d N\over\d\Rc}&=&\vc N_{\rm tot}{\d N\over\d L}=\vc\int\d
J_R\,f(L,J_R)\nonumber\\
&=&{\vc FN_{\rm
tot}\over\sigma^2}\int{\d\cE\over\Omega_R}\,\e^{-\cE/\sigma^2},
\end{eqnarray}
 where $N_{\rm tot}$ is the total number of stars in the disc.
 We adapt the technique described by \cite{DehnenDF} for determining $F(L)$
and $\sigma(L)$ from equations (\ref{givesVR}) and (\ref{givesSigma}) to the
determination of these quantities from equations  (\ref{givesVR}) and
(\ref{givesNL}). Specifically, we start from the values of  $F(L)$ and
$\sigma(L)$ that would hold in the epicycle approximation, when
$\Omega_R=\kappa$ independent of $\cE$ and
 \begin{equation}\label{givesF}
F(L)={\kappa\over\vc N_{\rm tot}}{\d N\over\d\Rc}.
\end{equation}
 Then at each $L$ we evaluate $\langle v_R^2\rangle$ from (\ref{givesVR}) and
multiply $\sigma$ by the ratio of the desired value to the value just
calculated. Then we re-evaluate $F$ from (\ref{givesNL}) and repeat until
convergence is obtained. 

We now address the question of how $\langle v_R^2\rangle$ should depend on
radius.  The scale heights $h$ of galactic discs are found to be largely
independent of radius \citep{vanderKruitS}, and for $h\ll R$ (when the
vertical dynamics can be considered one-dimensional) this finding implies
that the vertical velocity dispersion scales with the surface density as
$\Sigma^{1/2}$. If the ratio of the vertical and radial velocity dispersions
$\sigma_z/\langle v_R^2\rangle^{1/2}$ is independent of radius, as is often
assumed \citep[e.g.][]{KregelvdK}, then $\langle
v_R^2\rangle\propto\Sigma\propto\e^{-R/R_*}$. In the solar neighbourhood at
$R\simeq3R_*$ the oldest stars have $\sigma_R\gta40\kms$, so this line of
reasoning predicts that $\langle v_R^2\rangle^{1/2}\gta180\kms$ in the central regions,
which is implausibly large. 

Evidently these naive arguments based on complete decoupling of planar and
vertical motions are inadequate for the old disc; we need a distribution
function that treats the third integral properly. Pending the availability of
such a DF we have adopted the assumption that $\langle
v_R^2\rangle\propto\e^{-R/1.5R_*}$, which implies that at $R_*$ the old disc
has $\langle v_R^2\rangle^{1/2}\simeq85\kms$, which is only slightly lower
than the velocity dispersion in the Galactic bulge \citep{Richetal}.

From \cite{BinneySN} we take the time dependence of $\sqrt{\langle
v_R^2\rangle}(R_0)$ at solar galactocentric distance $R_0$ to be
 \begin{equation}\label{eq:sigR}
\sqrt{\langle v_R^2\rangle}(R_0,t)=
\max\left\{10,{38\left(\frac{t+0.038\Gyr}{10.038\Gyr}\right)^{0.33}}\right\}\kms,
\end{equation}
 which is consistent with the data of \cite{HolmbergNA}. 

\begin{figure}
\epsfig{file=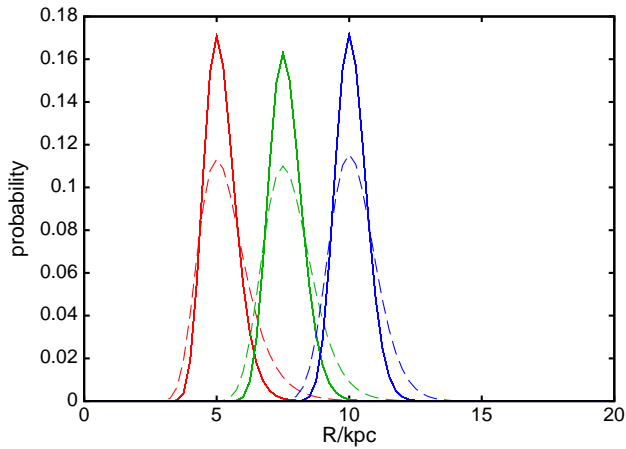, angle = -90,width=\hsize} \caption{The radial
distributions of stars with guiding centres at 5 (red), 7.6 (green) and
$10\kpc$ (blue) when at the Sun $\langle v_R^2\rangle^{1/2}=25\kms$ (full
curves) and $40\kms$ (dashed curves).
 \label{fig:blurr}}
\end{figure}

Fig.~\ref{fig:blurr} shows blurring distributions $P(R)$ from equation
(\ref{givesP}) for three radii ($5$, $7.6$ and $10\kpc)$ and two values of
$\langle v_R^2\rangle^{1/2}$ at the Sun, namely $25\kms$ and $40\kms$.

Note that scatterings by spiral arms and molecular clouds that heat the disc,
also change the angular momentum of each star and therefore its guiding
centre. Hence such scatterings contribute to both churning and blurring.

Since churning moves the guiding centres of the stars themselves we first
apply the churning matrix and apply the blurring matrix
afterwards.

\subsection{Vertical structure}

For comparison with observations of the solar neighbourhood we need to know
the vertical distribution of stars near the Sun. We determine this by
adopting a relationship between time and vertical velocity dispersion
\citep{BinneySN} 
 \begin{equation}\label{eq:sigz}
{\sigma_z}(\tau) = 
\max\left\{4,25\left(\frac{\tau}{10\Gyr}\right)^{0.33}\right\}\kms.
\end{equation}
 Further assuming that stars of a given age form an isothermal population,
their vertical density profile is
 \begin{equation}\label{eq:nofz}
n(z)\propto\e^{-\Phi(z)/\sigma_z^2},
\end{equation}
 where $\Phi(z)$ is the difference in the gravitational potential between
height $z$ and the plane. This potential is calculated for a model similar to
those presented by \cite{DehnenB} but with the thin and thick disc
scaleheights taken to be $0.3$ and $0.9\kpc$, the total stellar surface
density set to $35.5\msun\pc^{-2}$ with $3/4$ of the stellar mass in the thin
disc, and the gas surface density set to
$13.2\msun\pc^{-2}$ in conformity with \cite{Flynn06} and \cite{Juric08}.
The disc scalelength is taken to be $R_\d=2.5\kpc$ \citep{Robin03} and the
dark halo density is set such that $v_{\rm c}(R_0)=220\kms$.

\begin{table*}
\begin{tabular}{lllr}
Parameter&Meaning&Impact&Value\\
\hline
$\Sigma_{\rm crit}$&Kennicutt's threshold surface density&limited impact&0\\
$M_0$&initial gas mass&affects only $N(Z)$ at $\zeh<-0.7$; from Hess diagram&$3.0\times10^9\msun$\\
$M_1$&early infall mass&affects only $N(Z)$  at $\zeh<-0.7$; from Hess diagram&$4.5\times10^9\msun$\\
$M_2$&long timescale infall mass to $12\Gyr$&fixed by present mass&$2.9\times10^{10}\msun$\\
$b_1$&early infall timescale&affects only $N(Z)$ at $\zeh<-0.7$; from Hess diagram&$0.3\Gyr$\\
$b_2$&long infall timescale&limited impact; estimated from Hess diagram and
other work&$14\Gyr$\\
$f_{\rm A}$&Scheme A fraction of gas from IGM&free parameter&0.36\\
$f_{\rm B}$&Scheme B fraction of gas from IGM&free parameter fixed by local gradient&0.025\\
$k_{\rm ch}$&churning amplitude&free parameter&0.35\\
$t_0$&delay before first type Ia SNe&taken from literature&$0.15\Gyr$\\
$k^{-1}$&timescale for decay of type Ia SNe&taken from literature&$1.5\Gyr$\\
$f_{\rm eject}$&fraction of ejecta lost to Galaxy&small impact; mainly
affects metallicity scale&$0.15-0.04$\\ 
$f_{\rm direct}$&fraction of ejecta to cold ISM&small impact limited to
$\zeh<-0.7$&$0.01$\\
$t_{\rm cool}$&cooling time of warm gas&fixed by present mass of warm gas&$1.2\Gyr$\\
$M_{\rm warm}$&initial warm gas mass&impact limited to $N(Z)$ at $\zeh<-0.7$&$5\times10^8\msun$\\
$Z_{\rm IGM}$&metallicity of the IGM&limited to $R>12\kpc$; taken from literature&$0.1Z_\odot$\\
\hline
\end{tabular}
\caption{Parameters of the standard model. The infall rate is given by
equation (\ref{eq:Mdot}). The larger value of $f_{\rm eject}$ applies at
$R<3.5\kpc$. \label{tab:standard}}
\end{table*}

\begin{figure}
\epsfig{file=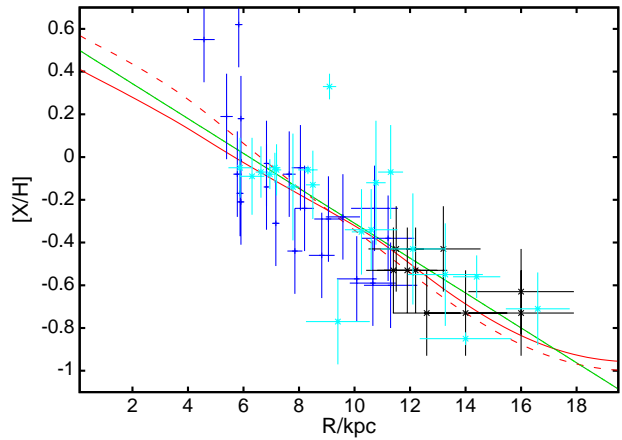,angle=-90,width=\hsize}
\caption{The metallicities of the current ISM in the standard model, [$Z$/H]
(red dashed curve) and [O/H] (red full curve), as functions of Galactocentric radius.
Also measurements of the metallicities of HII regions by Shaver et al.\
(1983) (dark blue), Vilchez \& Esteban (1996) (black crosses) and
Rolleston et al. (2000) (light blue crosses). The green line shows the linear least-squares fit to the measurements: it has a slope
of $-0.082\dex\kpc^{-1}$. The data points have been updated and rescaled to
$R_0=7.5\kpc$ as described in the text.
Where necessary points  have been shifted vertically by $ -8.93$ to put them on the solar scale.}
\label{fig:Zgrad}
\end{figure}

\section{The standard model}\label{sec:standard}

We now describe the properties of our standard model as a preliminary to
explaining how these properties depend on the input assumptions and the
values of the various parameters. In the standard model the accretion rate is
given by equation (\ref{eq:Mdot}); the values of the parameters for this
model are given in Table~\ref{tab:standard}.

\begin{figure}
\vbox{
\epsfig{file= 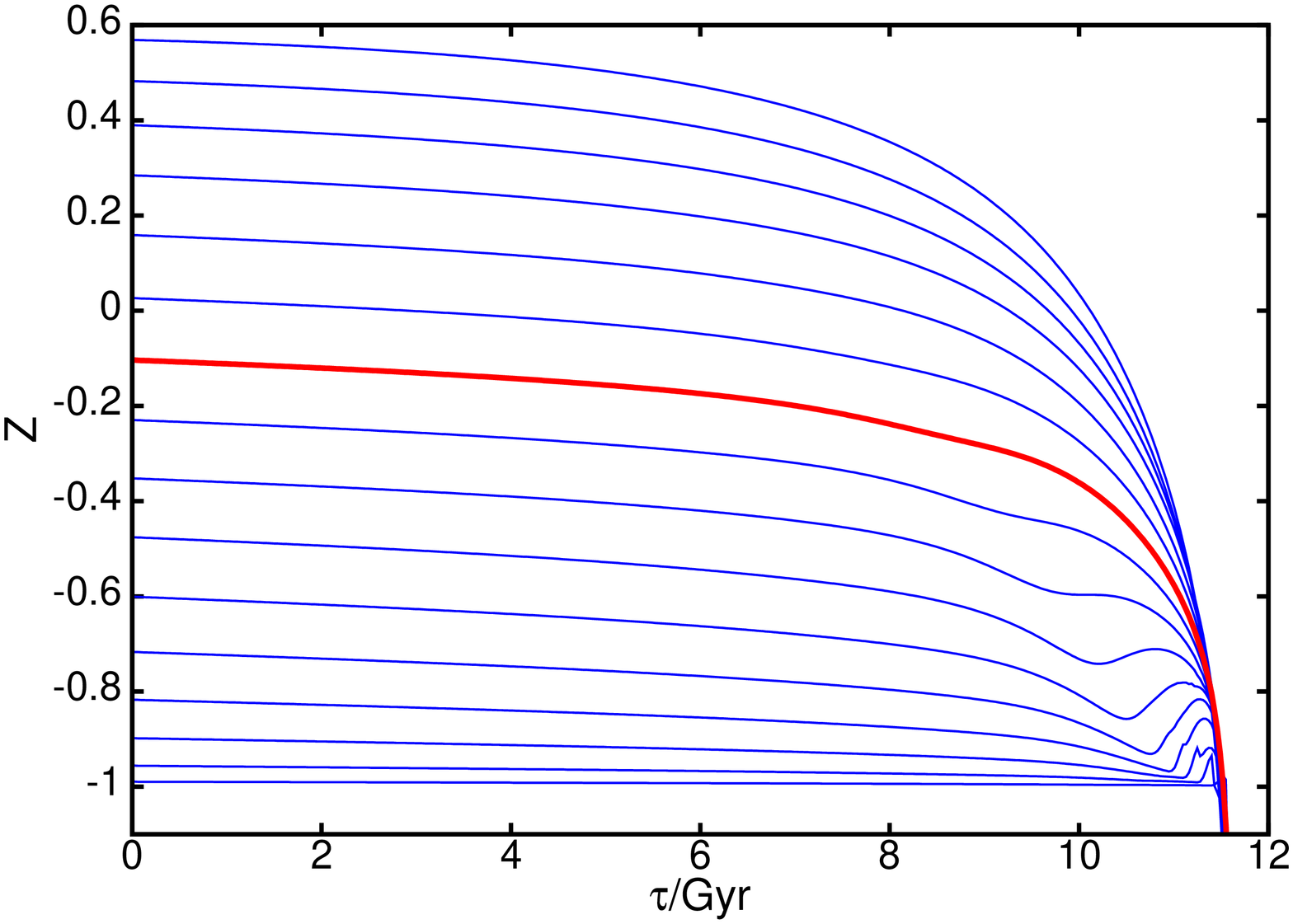,angle=-90,width=\hsize}\hfil
\epsfig{file= 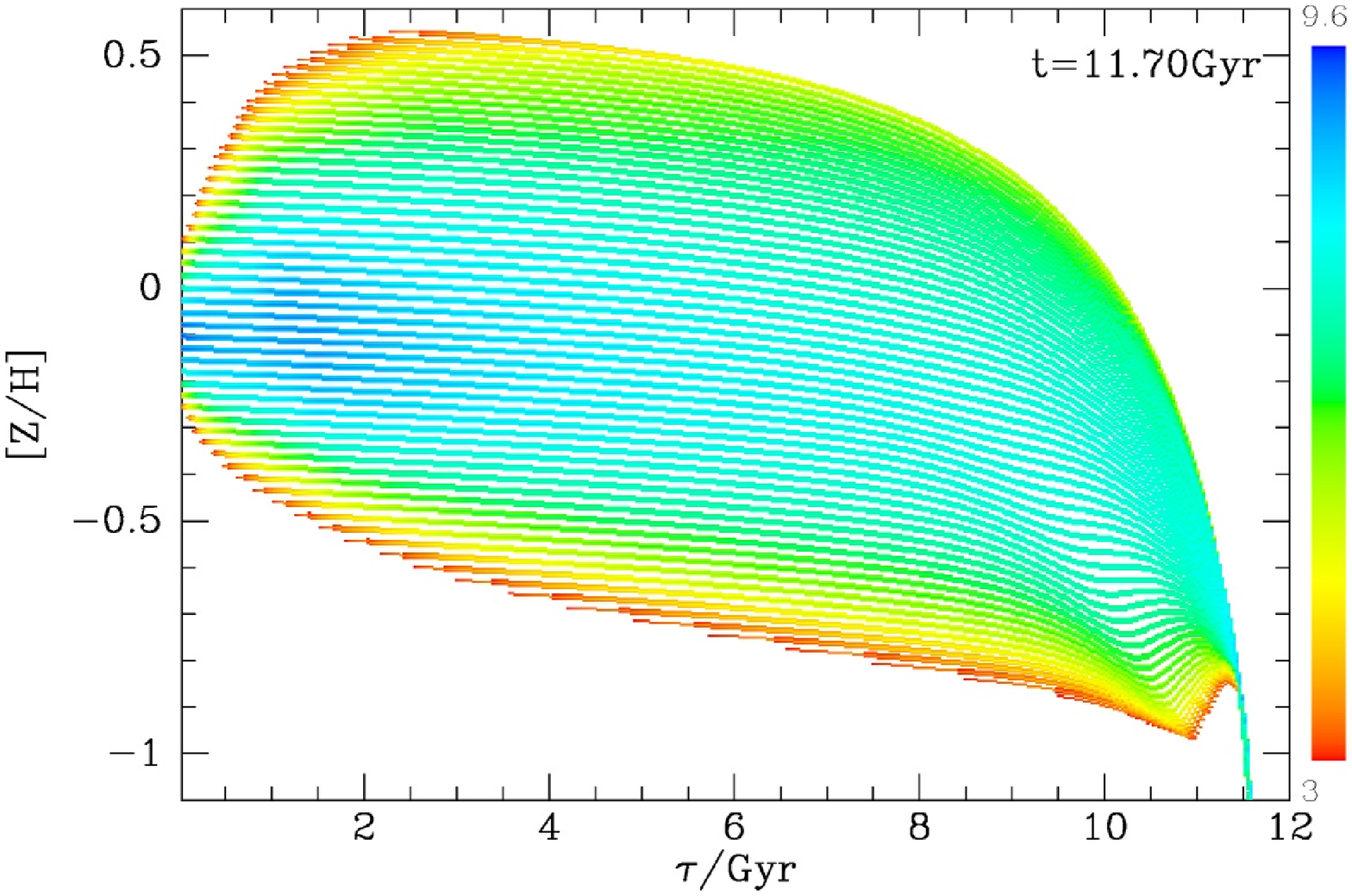,width=\hsize}
}
\caption{Upper panel: the
metallicity of the cold ISM in each annulus as a function of lookback time
showing each fifth ring.
The curve for the solar annulus is red. Lower panel: the present density of
solar-neighbourhood stars in the age-metallicity diagram. The colours encode
the logarithm of the density of stars.}
 \label{fig:ISMZ}
\end{figure}

The red dashed curve in \figref{fig:Zgrad} shows the current metallicity $Z$
of the ISM as a function of radius. There is quite a steep outward decline in
metallicity, the gradient in the vicinity of the Sun being of
order $-0.11\dex\kpc^{-1}$.
The solid red curve shows that [O/H] falls less
steeply with $R$ than does [$Z$/H], having a gradient near the Sun
$\sim-0.083\dex\kpc^{-1}$. The shallower gradient in oxygen reflects our use
of metallicity-dependent yields.  Although shallower gradients are generally
cited \citep[e.g][]{rolleston00} the data points in the figure are consistent with the model. The data
derive from \cite{shaver} who assumed $R_0=10\kpc$ and from \cite{vilchez}
and \cite{rolleston00},
who assumed $R_0=8.5\kpc$. To plot these data on a consistent scale with
$R_0=7.5\kpc$ we have when possible recalculated the Galactocentric distances
from the heliocentric distances, taking the latter from \cite{Kharch} or
\cite{Loktin} when possible. For some of the points in \cite{vilchez} and
\cite{rolleston00}
heliocentric distances were not available, so we simply reduced the
cited Galactocentric distance by $1\kpc$. The green line in
\figref{fig:Zgrad} is the 
linear least-squares fit to the data; its slope is $-0.082\dex\kpc^{-1}$.
Our gradient in [O/H] lies within the
range of frequently occurring values in Table 4 of \cite{Vila-CostasE92}, who
assembled data for 30 disc galaxies.

The upper panel in Fig.~\ref{fig:ISMZ} shows the evolution of $Z$ for the
cold ISM in a number of annuli -- the solar annulus is coloured red. The
smaller the radius of an annulus, the higher its curve lies in this plot
because chemical evolution proceeds fastest and furthest at small radii. At
small radii the metallicity of the cold ISM continues to increase throughout
the life of the Galaxy, whereas at $R\gta R_0\kpc$, $Z$ peaks at a time that
moves earlier and earlier as one moves out, and declines briefly before
flattening out.  This phenomenon reflects a combination of dilution by
infalling metal-poor gas and the inward advection of metals by the flow
through the disc. 

The lower panel in Fig.~\ref{fig:ISMZ} shows the corresponding present-day
metallicity distribution of solar-neighbourhood stars. Although this is the
distribution of stars currently in the solar annulus, it is clearly made up
of a series of curves, one for each annulus in the model. The curves for
interior annuli go from  green to yellow as one goes forward in time, reflecting
the fact that relatively recently formed stars are much less likely to have
moved a large radial distance than older stars. Similarly, in the bottom part
of the figure the colours go from blue to green to yellow as one moves towards
the time axis, because then one is moving over curves for larger and larger
radii, where both the star-formation rate and the probability of scattering
in to the solar radius are low. Hence, regardless of stellar age, most
solar-neighbourhood stars have $Z$ in a comparatively narrow range centred on
$[Z/\hbox{H}]\simeq-0.1$.

\begin{figure}
\epsfig{file=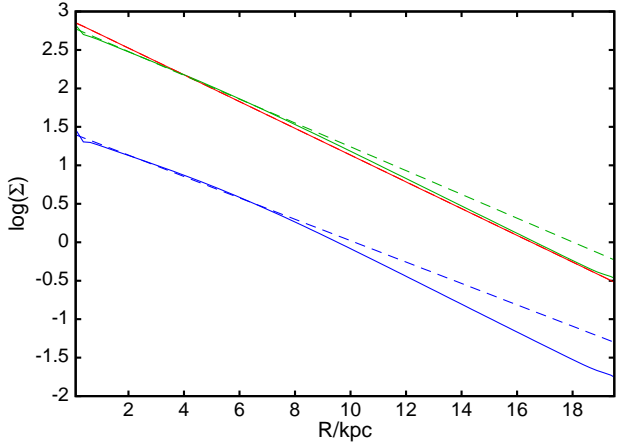,angle=-90,width=\hsize}
\caption{Full green curve: the surface density of the stellar disc at $11.7\Gyr$.
Broken green line: exponential fit to the inner part of this curve.  Red curve:
the surface density if stars remained where they were
born.  Blue curve: surface density contributed by stars born in the first
$0.8\Gyr$. Broken blue line: linear fit to this curve.} 
\label{fig:Sigstar}
\end{figure}

\begin{figure}
\epsfig{file=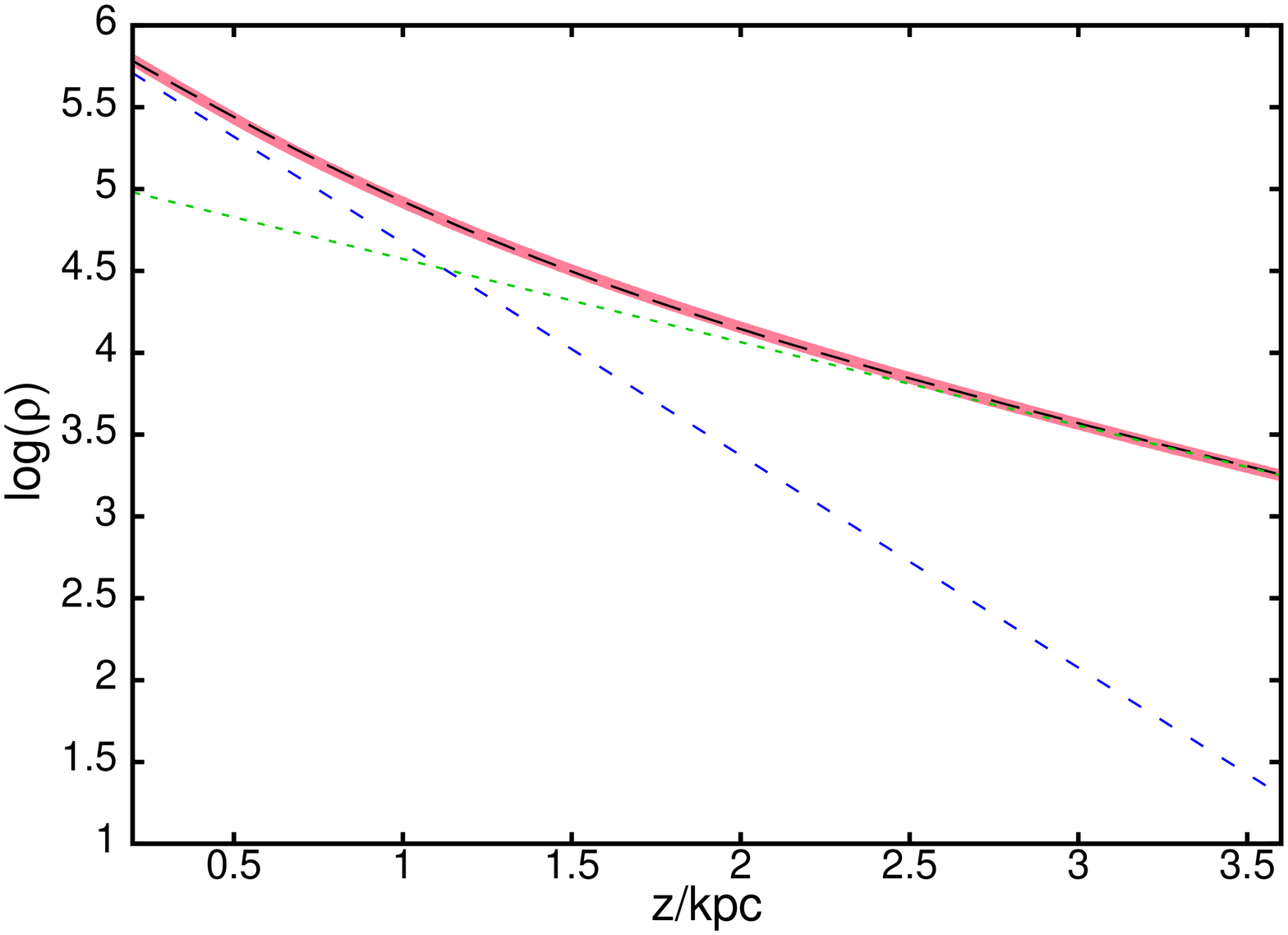,angle=-90,width=\hsize}
\epsfig{file=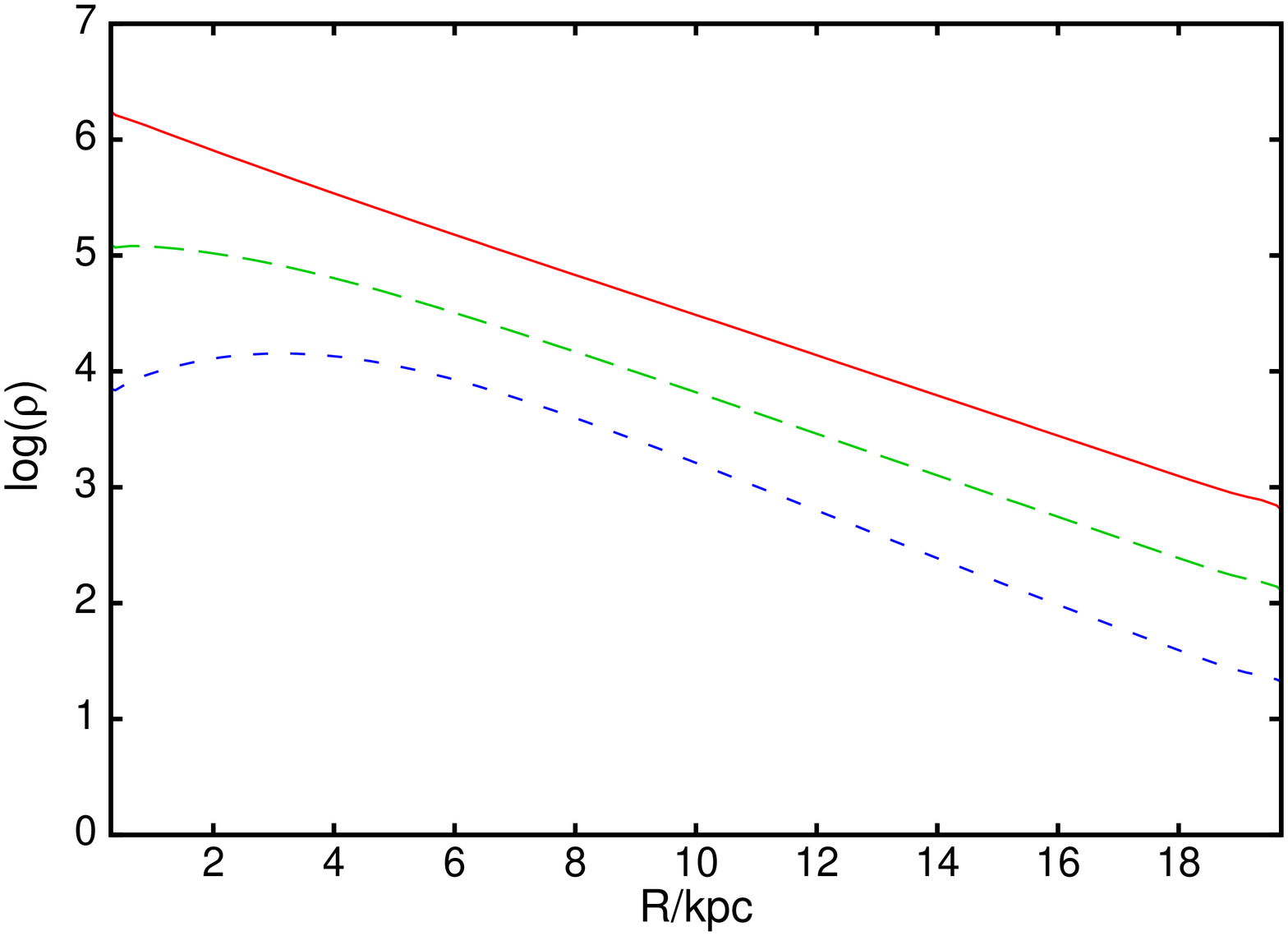,angle=-90,width=\hsize} \caption{Upper
panel: the volume density of stars at $R=7.6\kpc$ as a function of height
(red), a fit (black dashed) and its decomposition into  thin and thick
components. Lower panel:
the volume density of stars at the current epoch at $z=0$ (red), $z=0.75\kpc$
(green dashed) and $z=1.5\kpc$ (blue short dashed).
}
\label{fig:thick_thin}
\end{figure}

The full green curve in Fig.~\ref{fig:Sigstar} shows the surface density of the
stellar disc at $11.7\Gyr$, which is roughly exponential. The red points show
what the surface density would be if stars remained at their radii of birth.
By construction this forms an exponential disc with a scalelength of
$2.5\kpc$. The broken green line shows that at $R<R_0$ the disc approximates
an exponential with a larger scale length $\sim 2.8\kpc$. The blue curve shows
the surface density contributed by stars formed in the first $0.8\Gyr$, which
will be $\alpha$-enhanced. This distribution deviates more strongly from an
exponential because radial migration is most important for old stars. Fitting
an exponential to this curve at $R<R_0$ yields a scalelength $3.1\kpc$.

Fig.~\ref{fig:thick_thin} reveals that thin and thick disc components can be
identified within this overall envelope: the upper panel shows that the
vertical stellar density profile at the Sun is not exponential but can be
fitted by a sum of two exponentials. There is significant latitude in these
fits and the fraction of stars that is assigned to each component varies with
their scalelengths. For comparison with recent results of \cite{Juric08} we
present a fit with their value for the local thick disc fraction of $13$ per
cent. This yielded scaleheights of $h_1 = 335\pc$ and $h_2 = 853\pc$, very
well in the range of their results. Note that the double-exponential density
structure is {\it not\/} caused by any pecularity in star formation history,
like a peak in early star formation, but is a consequence radial mixing
combined with the given vertical force field. However, precise
characterisation of the vertical structure must await dynamical models that
employ a more accurate form of the third integral of galaxy dynamics.

The lower panel of
Fig.~\ref{fig:thick_thin} shows that at $z=1.5\kpc$ (where the thick disc is
dominant) the stellar distribution is less centrally concentrated than it is
in the plane; if one were to fit an exponential profile to the stellar
density at $z=1.5\kpc$ for $R<10\kpc$, the scalelength fitted would be larger
than that appropriate in the plane. Just this effect is evident in Fig.~16 of
\cite{Juric08}.

\begin{figure}
\epsfig{file=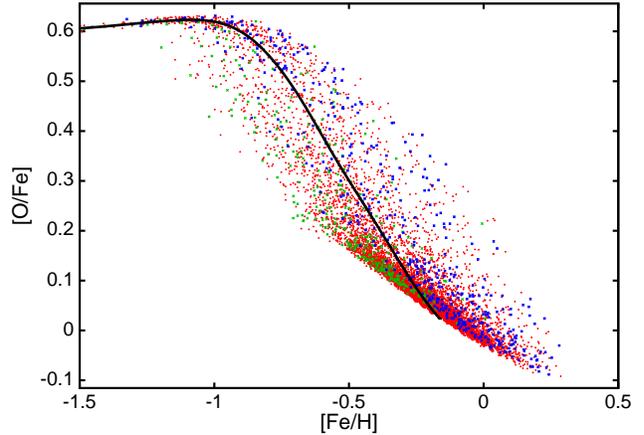,angle=-90,width=\hsize}
 \caption{The predicted distribution of solar-neighbourhood stars in the
([Fe/H],$\,$[O/Fe]) plane.  The sample is obtained by using the selection
function of the GCS survey as described in Section \ref{sec:Snhd} below. The
colours of points depend on the star's azimuthal velocity: $v_\phi<179\kms$
blue; $179<v_\phi/\!\kms<244$ red; $v_\phi>244\kms$ green.  The black curve
shows the trajectory of the solar annulus.}
 \label{fig:alpha_thick} 
\end{figure}

Fig.~\ref{fig:alpha_thick} shows the predicted distribution of
solar-neighbourhood stars in the ([O/Fe],$\,$[Fe/H]) plane when a sample is
assembled using the GCS selection function described in Section
\ref{sec:Snhd} below.  Two ridge-lines are evident: at top left of the figure
a population starts that stays at $\hbox{[O/Fe]}\simeq0.6$ until
$\hbox{[Fe/H]}\simeq-0.75$ and then turns down towards $(0,0)$, while a
second larger population starts at about $(-0.75,0.25)$ and falls towards
$(0.2,-0.05)$. This arrangement of points is very similar to that
seen in Fig.~2 of \cite{Venn04}. The upper ridge-line is associated with the thick
disc, and the lower ridge-line with the thin disc. In the appendix we show that
such bimodal distributions in [O/Fe] are a natural consequence of the
standard assumptions about star-formation rates and metal enrichment that we
have made. The structure is {\it not\/} a  product of the
double-exponential nature of the standard model's infall law; the model with
a constant gas mass displays exactly the same structure.  Breaks in the
Galaxy's star-formation history \citep{Chiappini97} and accretion events
\citep[][]{Bensby05} have been hypothesised to account for the dichotomy between the
thin and thick discs. Our models reproduce the dichotomy without a break or
other catastrophic event in our model's star-formation history. When
comparing \figref{fig:alpha_thick} with similar plots for observational
samples, it is important to bear in mind differences in  selection functions:
\figref{fig:alpha_thick} is for a kinematically unbiased sample, while most
similar observational plots are for samples that are kinematically biased in
favour of ``thick-disc'' stars.

The full curve in \figref{fig:alpha_thick} shows the trajectory of the
solar-neighbourhood ISM. At low [Fe/H] this runs along the ridge line of the
thick disc, and it finishes on the ridge line of the thin disc, but it is
distinct from both ridge lines. The sharp distinction between this curve and
the ridge line of the thin disc make it very clear that the latter is formed
through the migration of stars into the solar neighbourhood, {\it not\/}
through the chemical evolution of the solar neighbourhood itself. In many
previous studies it has been assumed that the ridge line of the thin disc
traces the historical evolution of the local ISM. \figref{fig:alpha_thick}
shows that this assumption could be wrong and that inferences
regarding the past infall and star-formation rates that are based on this
assumption are not to be trusted.

\begin{figure}
\epsfig{file=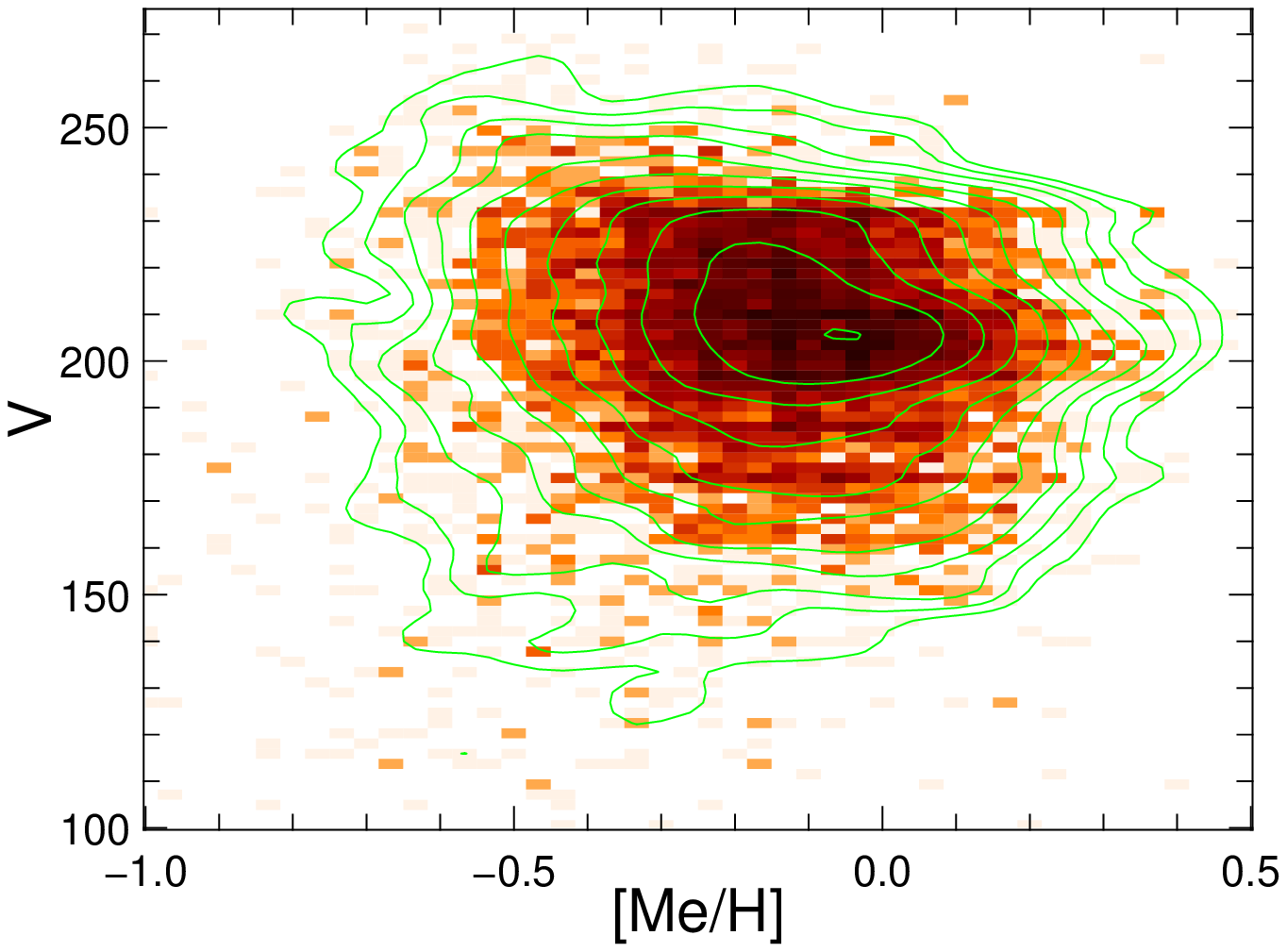,angle=0,width=\hsize}
\epsfig{file=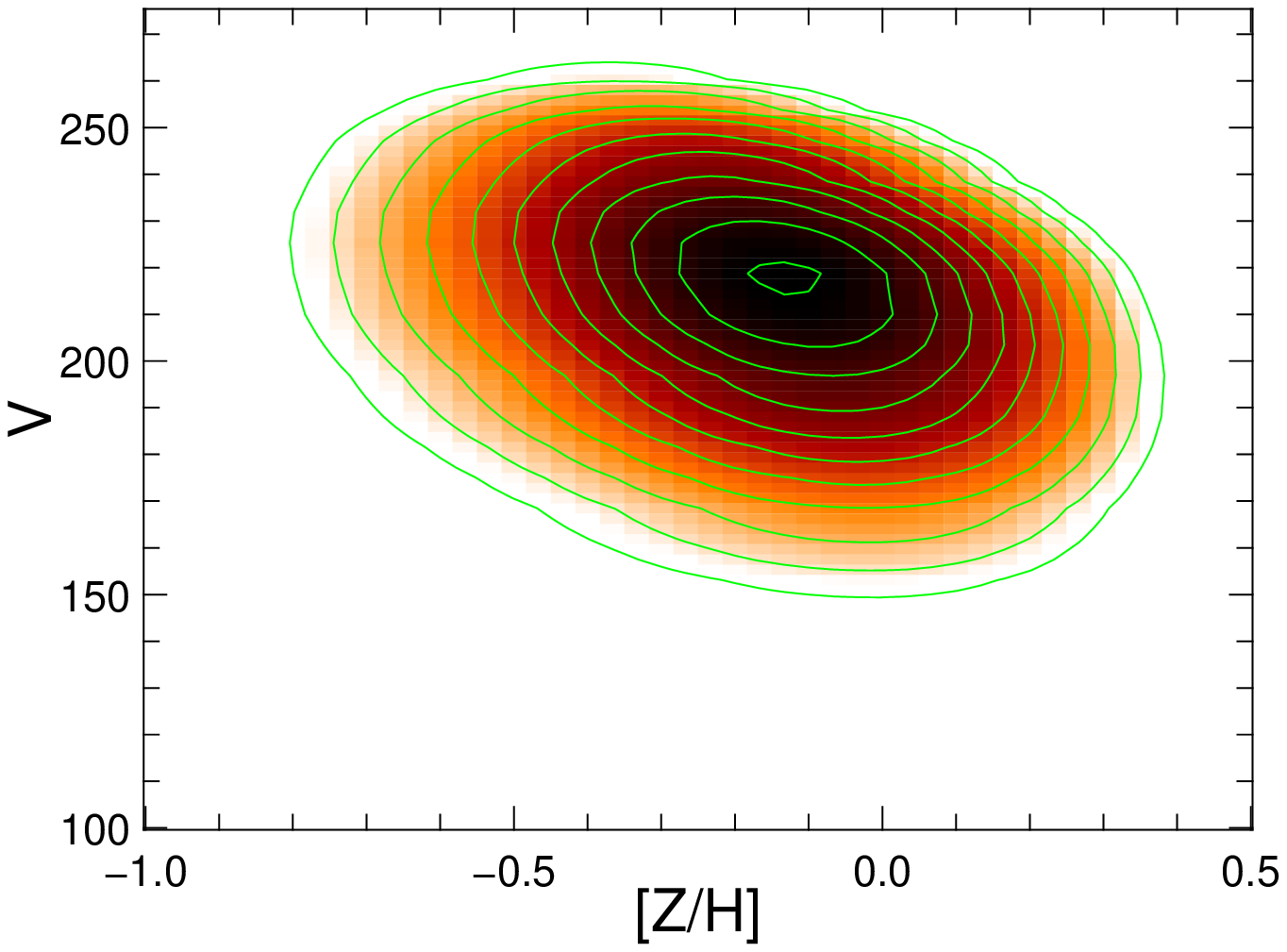,angle=0,width=\hsize}
\caption{Upper panel: the distribution of GCS stars in the
$([Z/\hbox{H}],v_\phi)$ plane. Lower panel: the prediction of the standard
model.Colours and contours reflect the density on a logarithmic scale with a
0.2 dex spacing for contours.}
\label{fig:vphiZ}
\end{figure}

In \figref{fig:alpha_thick} the points are colour coded by their angular
momenta/guiding centres: blue points are for $v_\phi<179\kms$ ($R_{\rm
g}<0.81R_0$), red points are for $179\kms\le v_\phi\le244\kms$ and green
points are for $v_\phi>244\kms$ ($R_{\rm g}>1.1R_0$).  At the low-metallicity
end of the thin-disc ridge line many points are green and few blue, while at
the high-metallicity end the reverse is true.  Thus low-metallicity thin-disc
stars tend to have guiding centres $R_{\rm g}>R_0$, while high metallicity
stars have $R_{\rm g}<R_0$.  \cite{Haywood08} has noted the same
metallicity-velocity correlations in samples of nearby stars. The thick disc
contains stars from all three radial ranges, but stars with small $R_{\rm g}$
(blue) are most prominent at higher [Fe/H].  

\figref{fig:vphiZ} shows the distribution of stars in the
([$Z$/H]$,\,v_\phi$) plane: the upper panel is for the GCS stars and the
lower panel is for the standard model. In both panels the highest density of
stars lies near $(0,220\kms)$ and the upper edge of the distribution rises as
one moves to lower metallicities. The metallicity gradient in the disc leads
to the main cluster of stars sloping downwards to the right.  A significant
difference between the two panels is that in the upper panel there are more
stars in the lower left region. This population is very much more prominent
in Fig.~5 of \cite{Haywood08}, where a band of points runs from small
$v_\phi$ and $\zeh$ up towards the main cluster.  This band is made up of halo
and thick-disc stars that are selected for in the samples from which Haywood
drew data. The other difference between Haywood's Fig.~5 and the lower panel
of \figref{fig:vphiZ} is that Haywood's main clump has a slightly less
pronounced slope down to the right.  It is likely that errors in the
measurements of [Fe/H] have moderated this slope. The GCS distribution shown
in the upper panel of \figref{fig:vphiZ} is (especially on the high
metallicity side) dominated by overdensities around the rotational velocities
of well-known stellar streams \citep[e.g., the Hercules stream][]{Dehnen98}.
This pattern overlays the general downwards slope. The model accounts well
for the steeper edge of the density distribution at high rotational
velocities, which is the combined effect of lower inwards blurring and lower
stellar densities from outer rings.

\begin{figure}
\epsfig{file=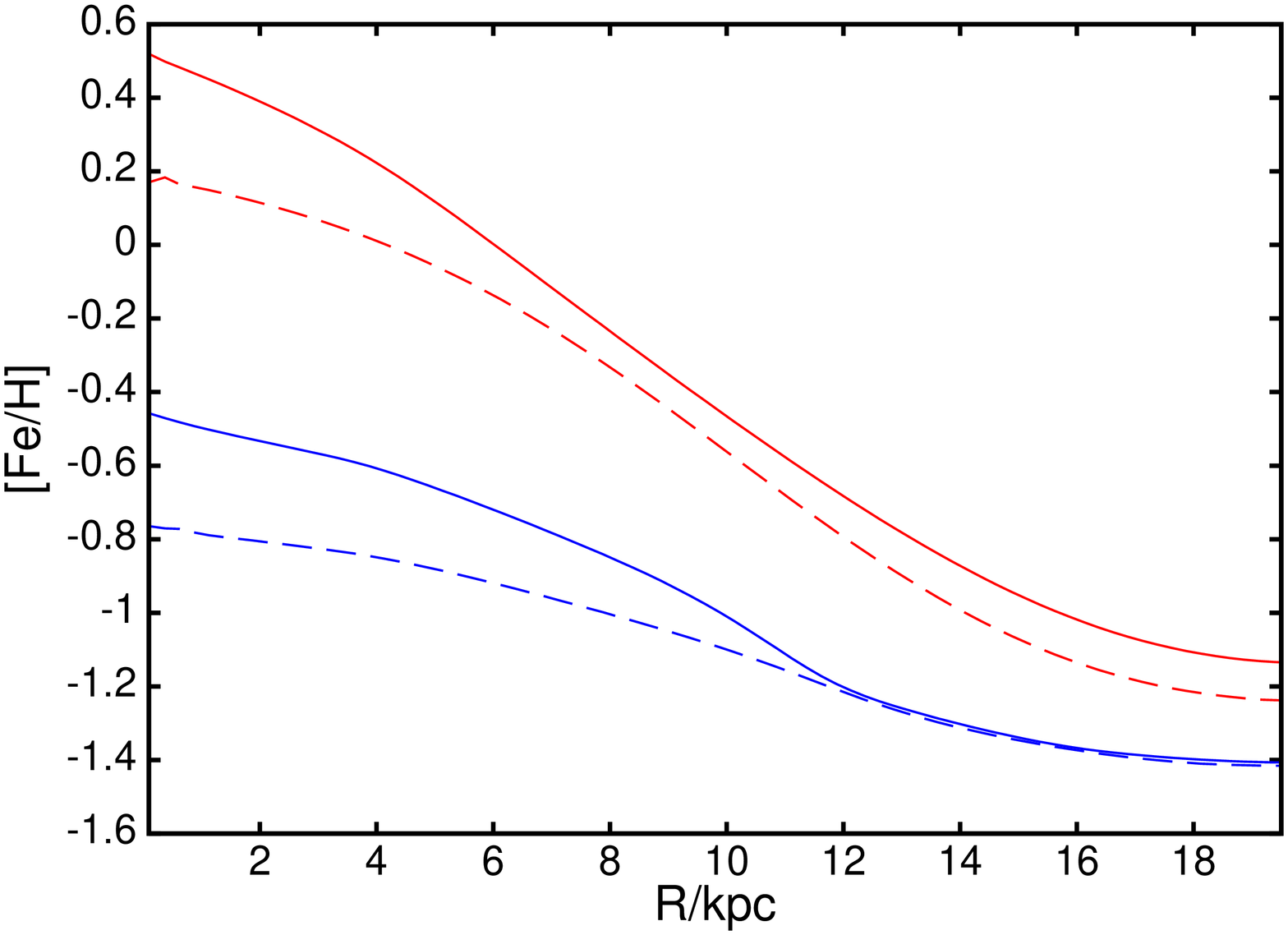,angle=-90,width=\hsize} 
\epsfig{file=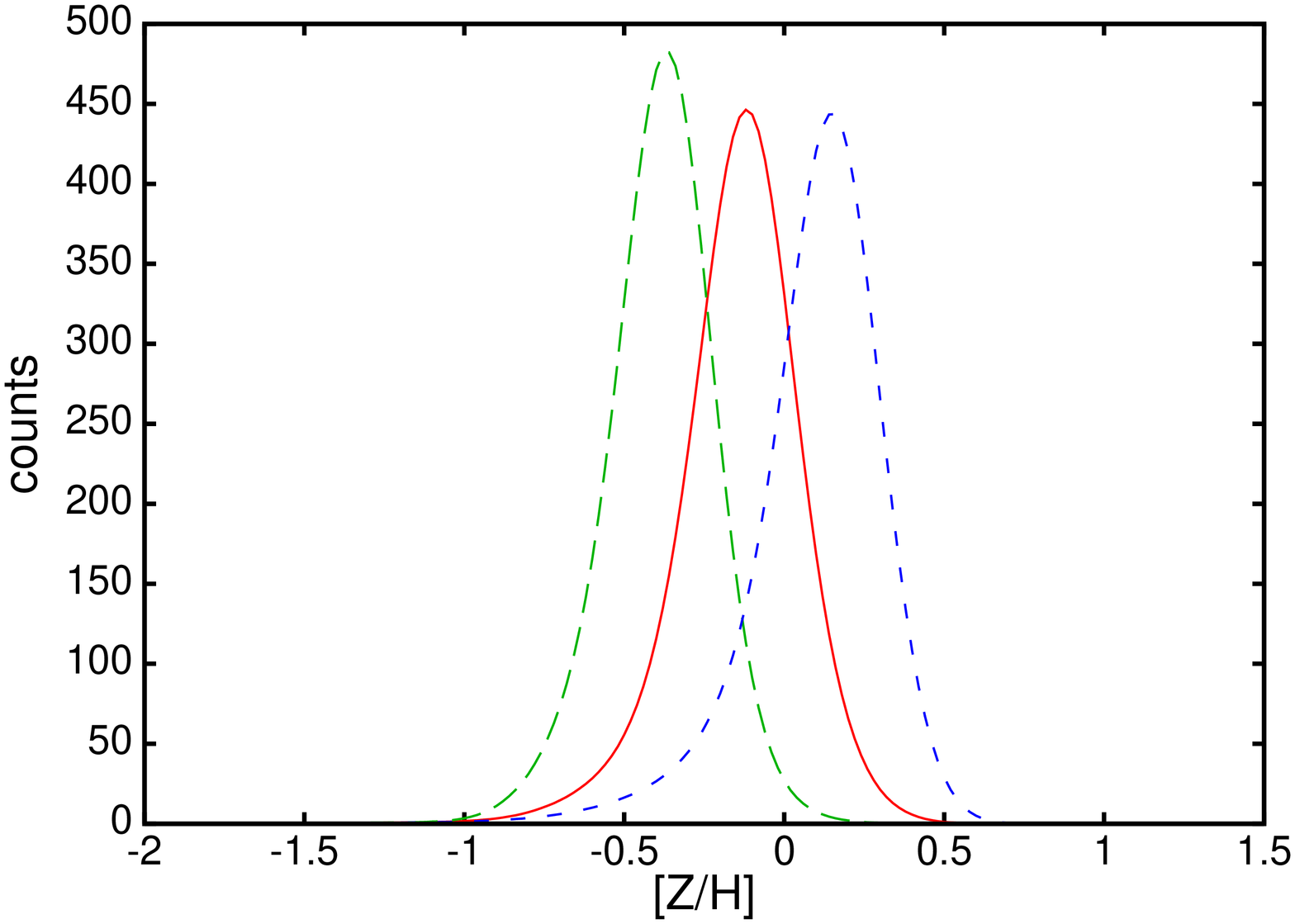,angle=-90,width=\hsize}
 \caption{Upper panel: mean metallicities of stars (dashed) and cold ISM
(full) as functions of $R$ at the present time (red) and at 1.5 Gyrs (blue).
Lower panel: the distributions over metallicity of stars currently at
$R=5\kpc$ (blue), $7.6\kpc$ (red), and $10\kpc$ (green). }
 \label{fig:star_gasZ}
\end{figure} 

The top panel of \figref{fig:star_gasZ} shows that the stellar metallicity
distribution is less centrally concentrated than that of the cold ISM from
which stars form.  Three factors are responsible for this result. First, the
mean metallicity of stars reflects the metallicity of the gas at earlier
times, which was lower. This effect is most pronounced at the centre, where
the metallicity of the ISM saturates later than further out. Second, radial
mixing, which flattens abundance gradients, has a bigger impact on stars than
gas because stars experience both churning and blurring. Third, the net
inflow of gas steepens the abundance gradient in the gas. 
\cite{HolmbergNA} have estimated the stellar metallicity gradient from the
GCS stars. When they select thin-disc stars they find 
$-0.09\dex\kpc^{-1}$, but when one excludes stars with $\zeh<-0.7$ (which 
ensures halo objects are removed), one obtains $-0.11\dex\kpc^{-1}$. The gradient of the dashed red line in
\figref{fig:star_gasZ} at $7.6\kpc$ is $0.10\dex\kpc^{-1}$ in excellent
agreement with the GCS data.

The lower panel of
\figref{fig:star_gasZ} shows the breadth of the metallicity distribution at
three radii. These distributions  have full-width at half 
maximum around $0.35\dex$ and are significantly offset to each other by
$0.25\dex$.

\begin{figure}
\epsfig{file=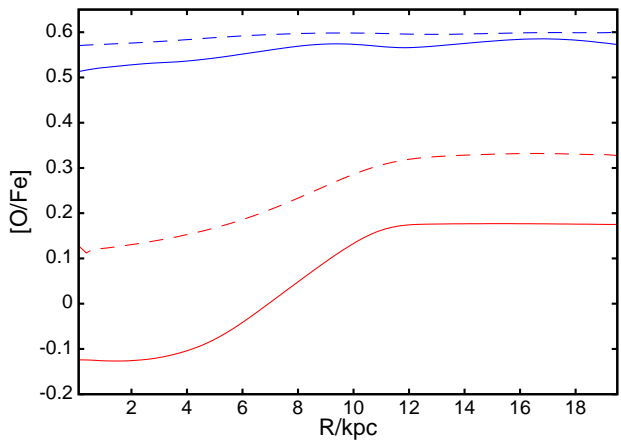,angle=-90,width=\hsize} 
 \caption{Full lines: [O/Fe] in the cold ISM after $1.5\Gyr$ (blue) and
$12\Gyr$ (red). Dashed lines: Mean [O/Fe] of stars after $1.5\Gyr$ and $12\Gyr$.}
 \label{fig:ISMalpha}
\end{figure}

Fig.~\ref{fig:ISMalpha} shows how $\alpha$-enhancement varies in time and
space, in stars and gas. Naturally, [O/Fe] declines with time in both the ISM
and in the stellar population, and at a given time is higher in the stars
than the gas.  [O/Fe] generally increases outwards but at $12\Gyr$ in both
stars and gas it attains a plateau at $R\gta10\kpc$, with $\alphaH\sim0.2$ in
the gas. The existence of the plateau is a consequence of the rule that in
the IGM [O/H] is the current value in the disc at $R\simeq12\kpc$; gas with
the given $\alpha$-enhancement rains on the disc at $R\lta20\kpc$, is
enriched by supernovae of both types and a few gigayears later arrives at
$R=12\kpc$ with its original $\alpha$-enhancement. This level is set by the
metallicity-dependent yields we have employed.

\begin{figure}
\epsfig{file=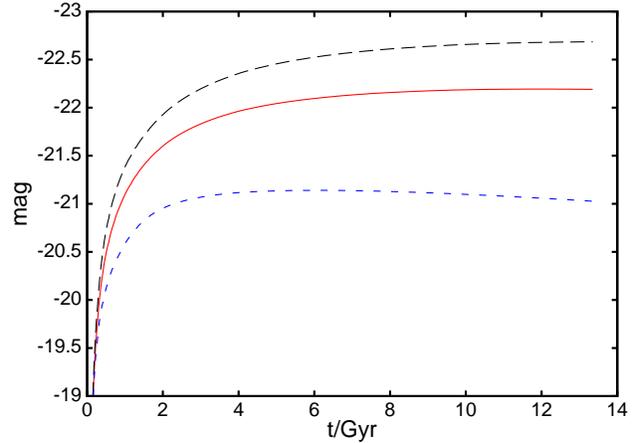, angle=-90, width=\hsize}
 \caption{Absolute magnitudes in the $B$ (blue short dashed), $R$ (red) and
   $I$ (black long dashed)
bands as functions of time.  No allowance has been made for obscuration.}
 \label{fig:Lt}
\end{figure}

\begin{figure}
\epsfig{file=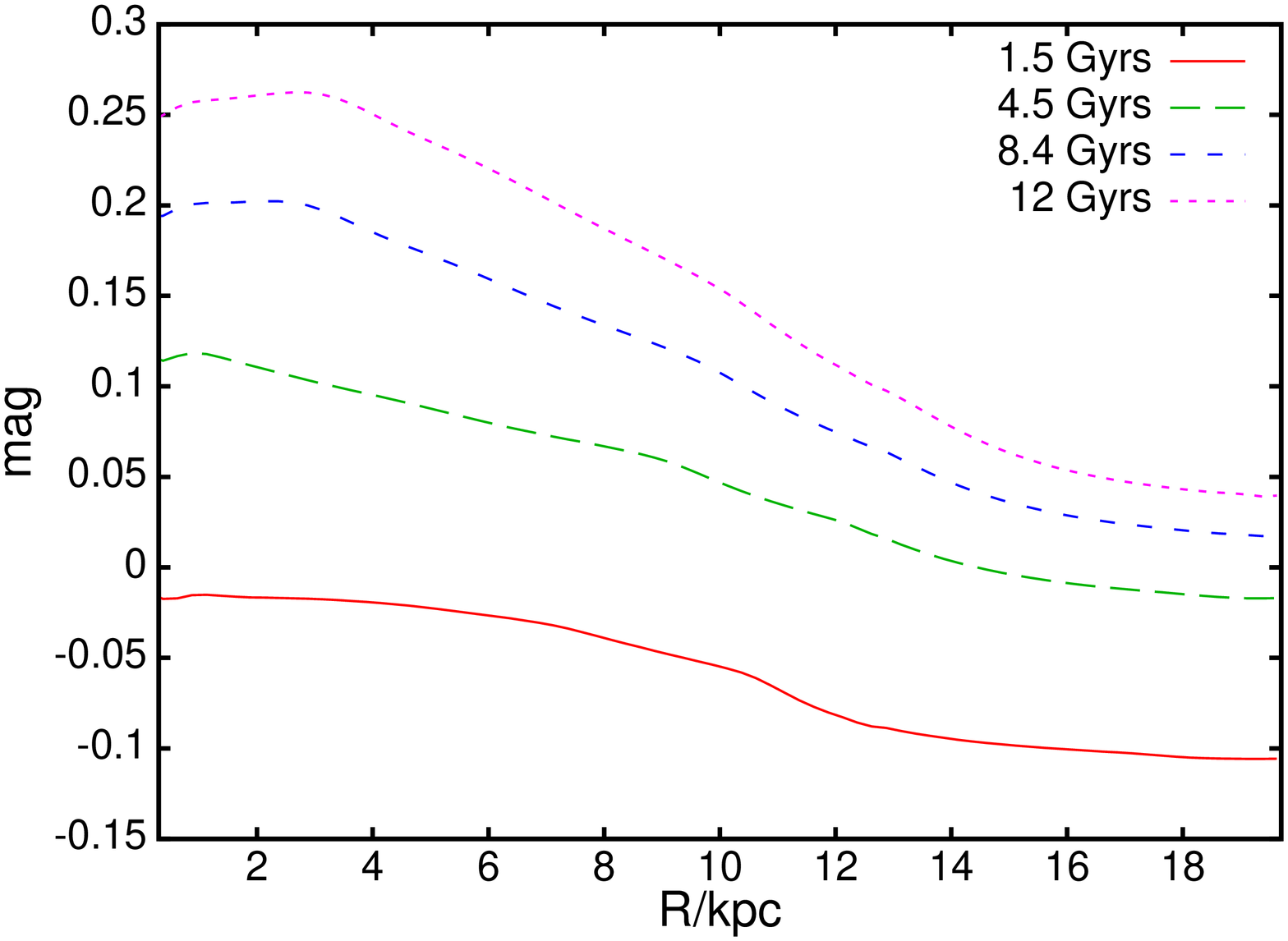,angle=-90,width=\hsize}
\epsfig{file=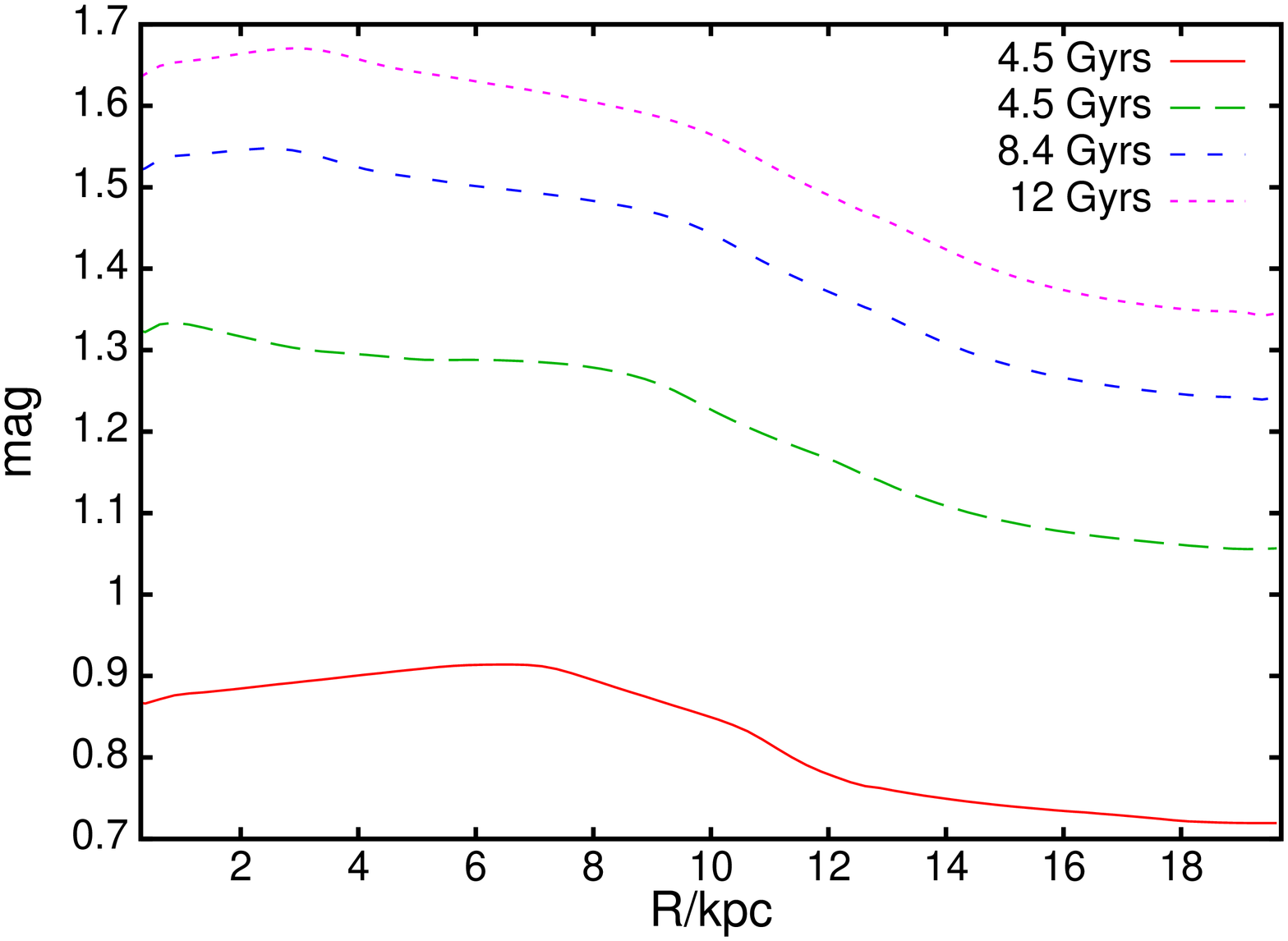,angle=-90,width=\hsize}
\caption{$U-B$ (upper panel) and $B-I$ (lower panel) as functions of radius
at $t=1.5$, 4.5, 8.4 and $12\Gyr$. No allowance has been made for
obscuration.}
\label{fig:colours}
\end{figure}

\figref{fig:Lt} shows the $B$, $R$ and $I$-band absolute magnitudes of the
standard model as functions of time. The $B$-band luminosity rises quickly
to a shallow peak around $5\Gyr$ and then commences a very slow decline. Emissions in the
$R$-Band are almost constant at the present time, while $I$-band luminosities continue to rise throughout the Galaxy's life because
additions to the stock of long-lived stars outweigh deaths of
relatively short-lived and predominantly blue stars. In our model the Galaxy
reaches an I-Band magnitude of around $-22.7$ which is exactly the
result one would expect for a disc galaxy with a rotation velocity of
$220\kms$ \citep[e.g.][]{Pizagno}

 \figref{fig:colours} shows the $U-B$
and $B-I$ colours of the disc at $t=1.5$, 4.5, 8.4 and $12\Gyr$. As expected,
the disc reddens at a declining rate throughout its life. There is at all times a
significant colour gradient between $R=8$ and $16\kpc$ that make the disc's
edge about $0.2\,$mag bluer in $B-I$ than its centre.

\begin{figure}
\epsfig{file=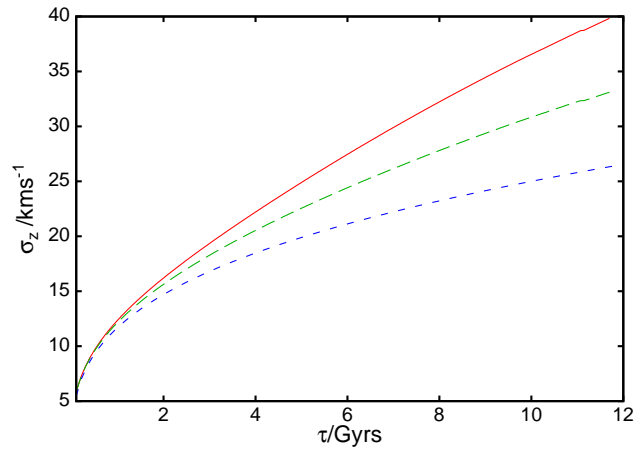,angle=-90,width=\hsize}
 \caption{Velocity dispersion of solar-neighbourhood stars as a function of
age. Red curve: $\sigma_z$ for all stars in the solar annulus. Green (long dashed) curve:
$\sigma_z$ for stars within $100\pc$ of the Sun. Blue (short dashed) curve: $\sigma_z$ for
stars born in the solar annulus.}
 \label{fig:sigmat}
\end{figure}

\figref{fig:sigmat} shows that radial migration causes the dependence of
velocity dispersion on time for stars that are currently in the solar
neighbourhood to differ materially from the acceleration law (\ref{eq:sigz})
that determines the time dependence of $\sigma_z$ for stars that are born at
given radius. The outward migration of stars brings to the solar
neighbourhood stars that carry with them the large velocity dispersions
characteristic of their places of birth. The impact that these migrants have
on $\sigma_z$ for stars of a given age increases with age, so at high ages
$\sigma_z$ increases faster than equation (\ref{eq:sigz}) predicts.
Least-squares fits of $\sigma_z\propto t^\beta$ to the red and green curves
in \figref{fig:sigmat} yield $\beta=0.53$ and $0.44$, respectively.
Empirically the photometrically complete portion of the Hipparcos catalogue
shows that the best power-law fits to the rate of increase of $\sigma_z$
yield $\beta\simeq0.45$ \citep{JustJahreiss07,AumerB08}. From a theoretical
standpoint, this result has hitherto been puzzling because the largest
exponent that can be obtained from the dynamics of star scattering is $1/3$
\citep{BinneyL88}. Such studies treat the acceleration as a local process.
Our result suggests that the conflict between theory and observation is
attributable to violation of this assumption.

\section{Fitting the model to the solar neighbourhood}\label{sec:Snhd}

A major constraint on the models is provided by comparing
the model's predictions with samples of stars observed
near the Sun. To make these comparisons, we have to reproduce the selection
functions of such samples, which proves a non-trivial job.

The GCS is an important sample, and for each model we calculate the
likelihood of this sample.  \cite{Nordstrom04} obtained Str\"omgren
photometry and radial velocities for a magnitude-limited sample of $16\,682$
F and G dwarfs, nearly all of which have good Hipparcos parallaxes.  From the
photometry they estimated metallicities and ages. There has been some debate
about the calibration of the metallicities and ages
\citep{Haywood06,HolmbergNA,Haywood08}. Recently the re-calibrated data from
\cite{HolmbergNA} became available and it is to these data that we have
compared our models. We compare their metallicities ($\meh$) to our $\zeh$ as it is not 
entirely obvious to what extent alpha enrichment enters into their measurements. 
Since assigning ages to individual stars is very
difficult, we have concentrated on matching the distribution of stars in the
$(M_V,T_{\rm eff},Z)$ space from which ages are derived.

For each metallicity we construct a volume-limited stellar number density of
stars in the $(M_V,T_{\rm eff})$ plane by considering each annulus $j$, and
calculating the fraction of each population in this annulus that will be in
the solar neighbourhood.  For given absolute magnitude, the probability that
a star will enter the sample is
 \begin{equation}
W(r_{\rm max})=\int_0^{r_{\rm max}}\d
r\,r^2\int\d^2\Omega\,n(z),
\end{equation}
 where the space density of stars $n(z)$ is assumed to be plane parallel and
given by equation (\ref{eq:nofz}).  

For the GCS selection function we use the approximate $b-y$ colour rules from
\cite{Nordstrom04} -- a more sophisticated selection function could in
principle be constructed, but it is not possible from the published data. At each colour, the appropriate selection
function $\phi$ is characterised by two apparent magnitudes $v_1$ and $v_2$
listed in Table \ref{tab:selfn}: $\phi$ declines linearly from unity at
magnitudes brighter than $v_1$ to zero fainter than $v_2$. $\phi$ vanishes
for $b-y$ bluer than $0.21$. For $b-y$ redder than $0.38$, $\phi$ is reduced
by a factor $0.6$.

The selection of \cite{Nordstrom04} is designed to exclude red giant 
branch (RGB) stars from the data set. However, some of these stars are still in
the sample, while the sample is biased against stars just below the giant
branch. We take out of consideration
the RGB itself and we downscale the theoretically
expected population density near the starting point of the red giant branch
by a factor of 4 to reconcile it with the data. Since
the number of RGB-stars in the GCS is not large anyway, the loss of
information is small. We also removed from the dataset three objects that
are far too faint to be attributed to the main sequence. The theoretical
distributions are convolved with a Gaussian of dispersion $0.1\dex$ in
[$Z$/H] to allow for measurement errors.

\begin{table}
\caption{Magnitudes defining the Geneva--Copenhagen  selection
function\label{tab:selfn}}
\begin{tabular}{c|ccccccc}
$b-y$&$0.21-0.25$&$<0.344$&$<0.38$&$<0.42$&$>0.42$\\
\hline
$v_1$&7.7& 7.8 & 7.8 & 7.8 & 8.2\\
$v_2$&8.9& 8.9 & 9.3 & 9.3 & 9.9\\
\end{tabular}
\end{table}

\begin{figure}
\epsfig{file=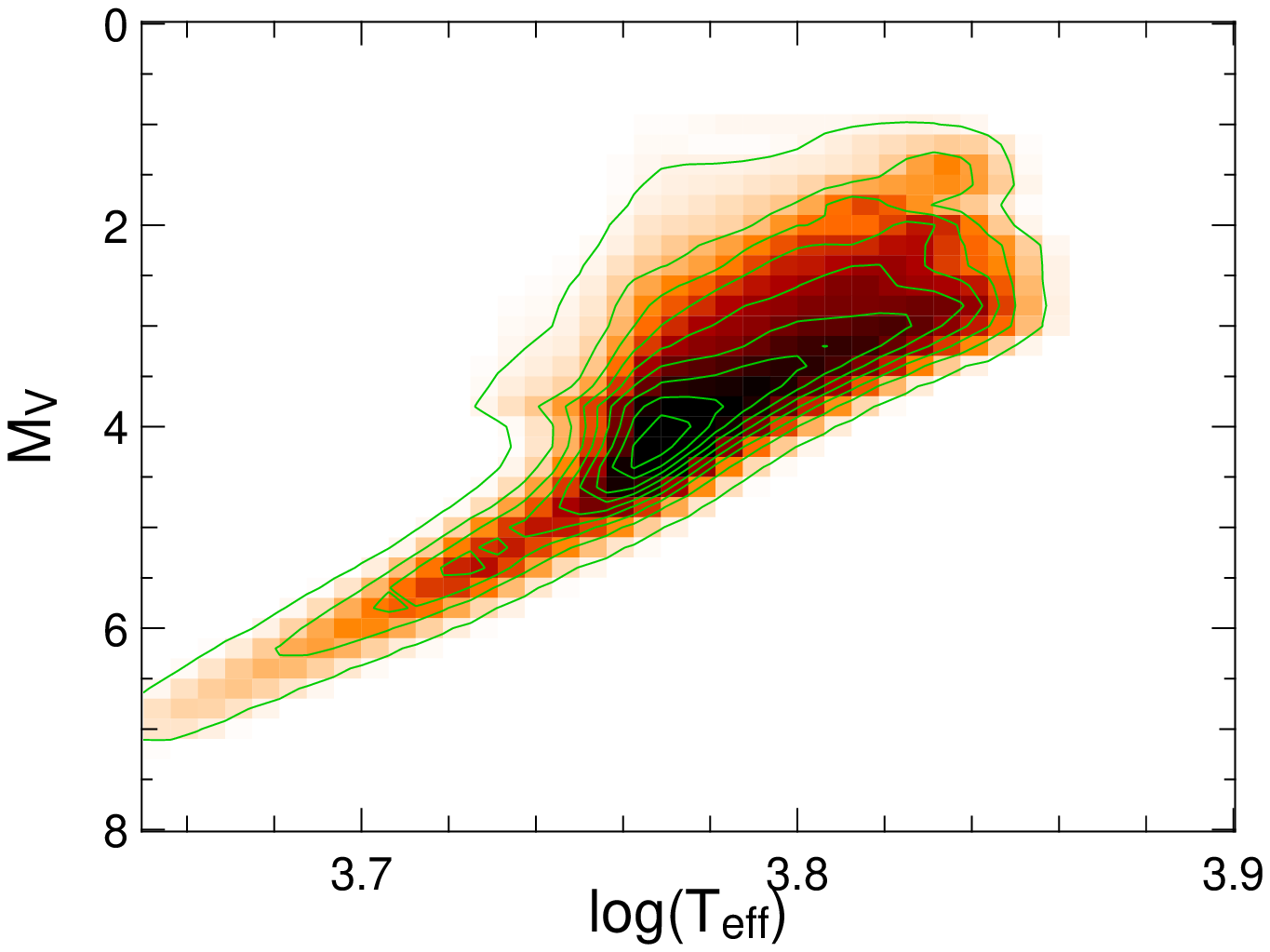,angle=0,width=\hsize} 
\epsfig{file=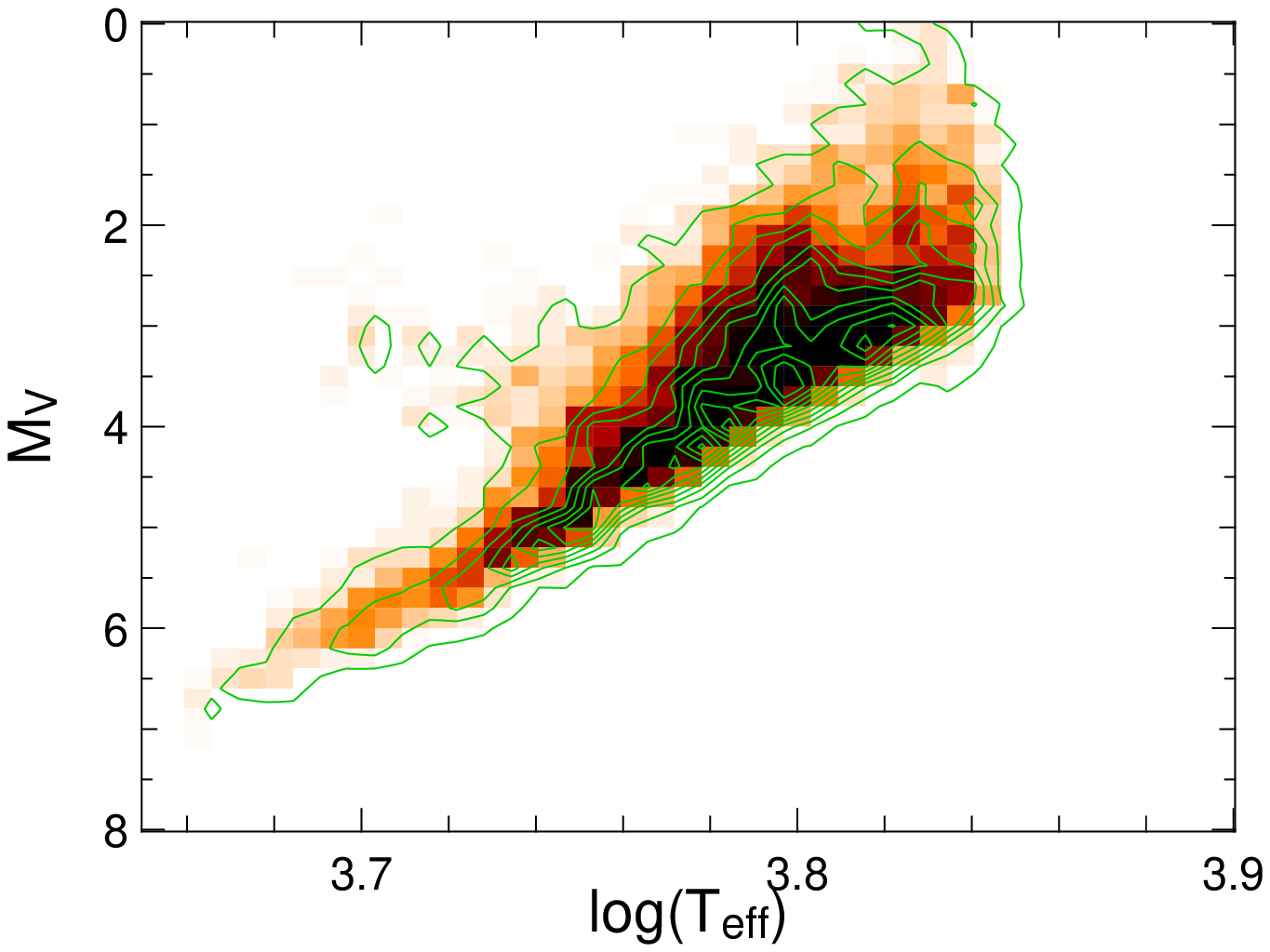,angle=0,width=\hsize} 
\caption{Comparison of the observed (upper) and predicted (lower)
distributions of GCS stars in the $(T_{\rm eff}, M_V)$ plane. Contours have an
equal spacing of 10 counts per bin, starting with 3.
\label{fig:GCScomp}}
\end{figure}

Fig.~\ref{fig:GCScomp} compares predicted (lower panel) and observed (upper
panel) Hess diagrams for the GCS stars.  As discussed by \cite{HolmbergNA},
the ridge-line of the main sequence in the GCS data is significantly
displaced from that predicted by isochrones. We have eliminated the effects
of this offset on \figref{fig:GCScomp} in the simplest possible way, namely
by decreasing all model values of $\log(T_{\rm eff})$ by the value, $0.015$,
that yields the closest agreement between the theoretical and observation
main sequences.  After this correction has been made, the agreement between
the theoretical and observational Hess diagrams shown in \figref{fig:GCScomp}
is convincing though not perfect. The original conception had been to
determine the model's parameters by maximising the likelihood of the GCS
stars in the model density in $(M_V,T_{\rm eff},Z)$ space, but confidence in
this plan was undermined by (i) the need for an arbitrary alignment of
measured and theoretical values of $T_{\rm eff}$, and (ii) the extent to which
the likelihood of the data depends on the uncertain GCS selection function.
Notwithstanding these reservations, we are encouraged that the standard model
maximises the likelihood of the data at an age, $\sim11\Gyr$, that agrees with
other estimates of the age of the solar neighbourhood \citep[][ and references
therein]{AumerB08}.

\begin{figure}
\epsfig{file=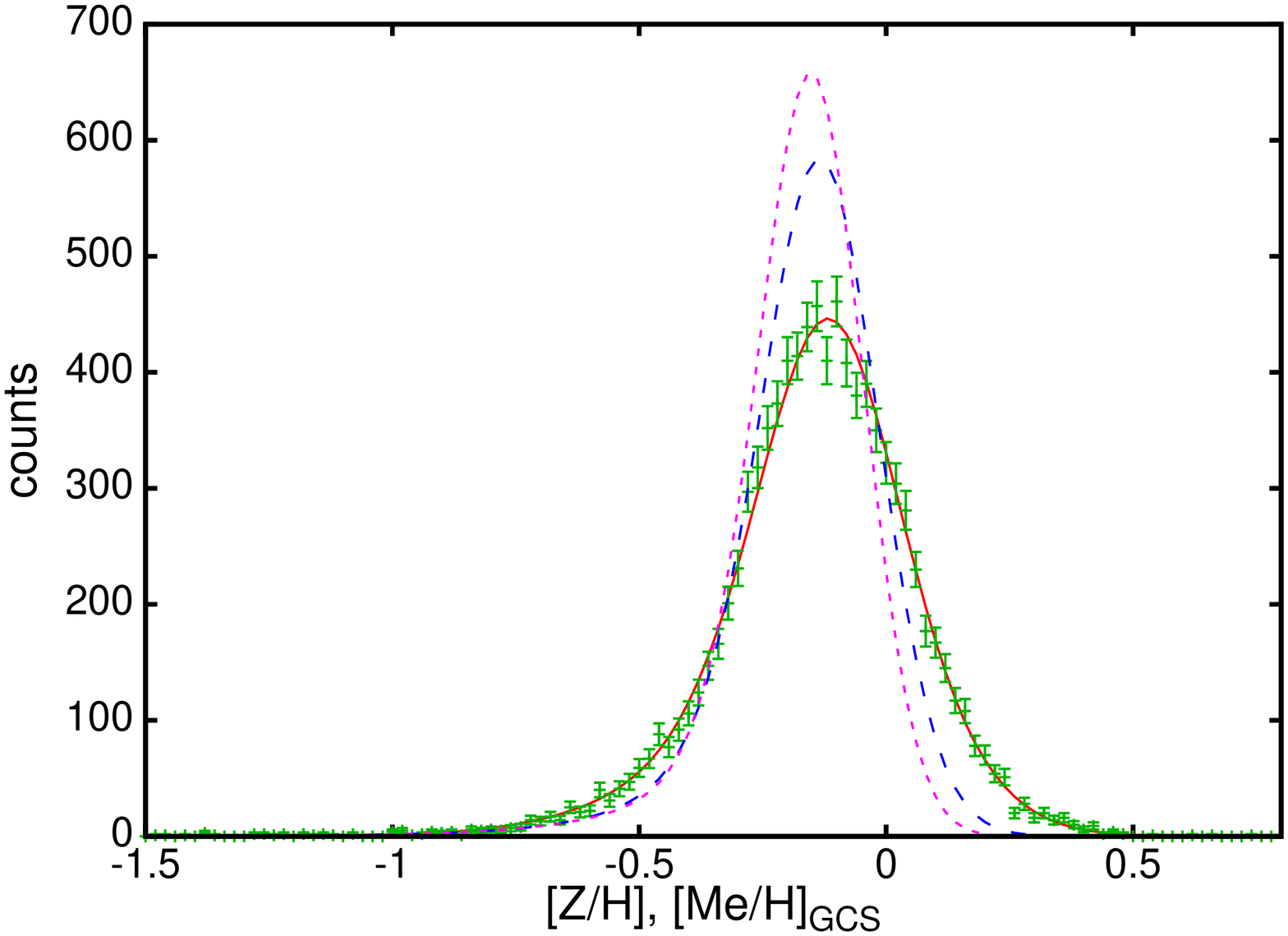,angle=-90,width=\hsize}
\epsfig{file=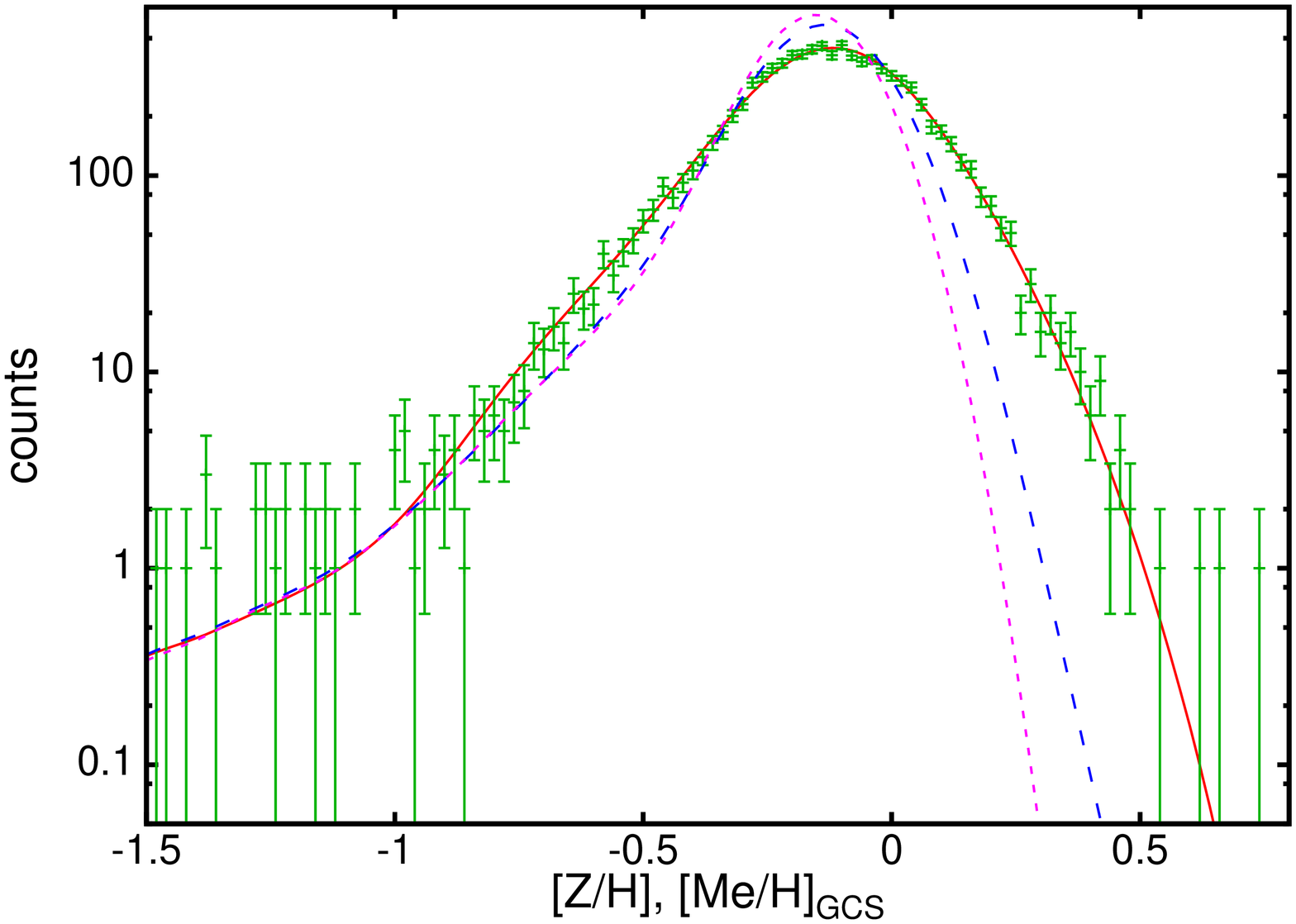,angle=-90,width=\hsize}
 \caption{The metallicity distribution of GCS stars (green points) and the
corresponding prediction of the standard model (full red curve). The broken blue curve
shows the model that differs from the standard model only in the elimination
of churning and radial gas flows. The broken pink curve shows the model with
neither churning nor blurring.  The lower panel shows the same data but with
a logarithmic vertical scale to reveal structure in the wings of the
distribution.}
 \label{fig:metdis}
\end{figure}

The full red curve in \figref{fig:metdis}  shows the metallicity
distribution of stars predicted by the standard model, and green points show
the GCS data.  The agreement is excellent.  At the lowest
metallicities theory predicts slightly  too few stars, but the uncertainties in both
the theory and the data are large in this limit and the theory
does not include halo stars, so it should under-predict the data at
$[Z/\hbox{H}]\lta-1$.

\section{General trends}\label{sec:trend}

We now discuss aspects of how the observable properties of a model depend on
its parameters, and what is required to achieve fits of the quality seen in
Figs~\ref{fig:Zgrad} and \ref{fig:metdis}.

Table~\ref{tab:standard} lists the model's 16 parameters; the third column
explains whether the parameter is fitted to the GCS data or taken from the
literature, and indicates the sensitivity of models to that parameter. Five
parameters ($M_0, M_1, b_1,f_{\rm direct},M_{\rm warm}$) significantly affect
only the distribution of stars at $\zeh<-0.7$. Six other parameters ($M_2,
t_0,k^{-1},t_{\rm cool},Z_{\rm IGM}$) are fixed by observed properties of the
Galaxy other than the solar-neighbourhood stellar distribution. The remaining
five parameters are the critical surface density for star formation
$\Sigma_{\rm crit}$, the long infall timescale $b_2$, the accretion
parameters $f_{\rm A}$ and $f_{\rm B}$ and the churning strength $k_{\rm
ch}$. We shall see that $f_{\rm B}$ is effectively set by the metallicity
gradient in the gas, that $b_2$ is effective determined by the local Hess
diagram, and that the value of $\Sigma_{\rm crit}$ is unimportant providing it
is small (we have included this parameter only for consistency with earlier
work; the models do not want it). Consequently the fit of the model to the
data shown by the red curve and the green points in \ref{fig:metdis} is
obtained by adjusting just $f_{\rm A}$ and $k_{\rm ch}$.

The number of stars more metal poor than $\zeh\sim-1$ depends
sensitively on the thermal structure of the early ISM. Most previous studies
\citep[exceptions include][]{Thomas98,SamlandG} have used only one phase of
the ISM. Introducing the warm component of the ISM delays the transfer of
metals to the star-forming cold ISM by $\sim1\Gyr$, thus increasing the
number of extremely metal-poor G dwarfs. Our first models initially had no
warm gas, with the result that at early times the mass of warm gas was
proportional to time and the metallicity of the cold gas rose quadratically
with time. These models had an over-abundance of {\it very} metal-poor stars.
These experiments led to the conclusion that pregalactic and halo stars 
endowed the disc with
warm, metal-rich gas at the outset. Even at late times, the existence of the
warm ISM delays the introduction of freshly-made metals into stars, and thus
in concert with the gas flow through the disc steepens the metallicity
gradient in the stellar disc; eliminating  the warm component raises the
metallicity of the solar neighbourhood and beyond by $\sim0.1\dex$.

The metallicity of infalling gas only affects the structure of the disc at
$R\gta 12\kpc$. Lowering $Z_{\rm IGM}$ steepens the metallicity gradient at
$R>R_0$.

It is instructive to consider the case in which blurring is included but
churning is turned off by setting $k_{\rm ch}=0$ and radial flows are
eliminated by ensuring that $f_{\rm A}+f_{\rm B}=1$ so every ring's need is
fully supplied from the IGM. The broken blue curve in \figref{fig:metdis} shows the
present-day metallicity distribution that this model predicts for the GCS.
The peak of the distribution is much narrower than in the standard model, and
there is a striking deficiency of metal-rich stars. The broken pink curve
shows the effect of also turning off blurring: the deficiency of metal-rich
stars becomes even more striking but there is negligible change on the
metal-poor side of the peak.

Reducing the current SFR by making the infall rate a more rapidly declining
function of time shifts the peak of the distribution to higher metallicities
by reducing the relative strength of recent inflow and thus the supply of
fresh, metal-poor gas.  The use of an IMF that is steeper in the low-mass
region, as has been suggested by some studies, reduces the mass of metals
that is locked up in low-mass stars and increases the metal production by
each generation. The metallicity in such a regime is accordingly higher, but
the shape of the distribution does not change.  Since changes in the IMF at
the high-mass end, which is equally uncertain, can produce compensating
variations in yields, loss rates, etc., we stayed with the traditional
approach via the Salpeter IMF.  

In all interesting models the metallicity of the local ISM saturates early
on. The saturation level depends on the pattern of gas flow through the disc,
and on the current SFR relative to the mean rate in the past: the faster the
decline in the SFR, the higher the current metallicity.  Naturally the stars
of the solar neighbourhood are on the average younger in models with a
constant gas mass than in models in which the infall rate is declining
according to equation (\ref{eq:Mdot}). This relative youth is reflected in
the structure of the local Hess diagram. We reject the model with constant
gas mass because it assigns a significantly smaller likelihood to the Hess
diagram of the GCS stars than does a model based on equation (\ref{eq:Mdot}).

\subsection{Selecting the standard model}

All our models have quite strong metallicity gradients in both stars and gas
(Figs \ref{fig:Zgrad} and \ref{fig:star_gasZ}).  Since the metallicity of the
central gas is enhanced by radial gas flow, and models with large $f_{\rm B}$
have larger central flows than models with large $f_{\rm A}$ and vice versa
at large radii (\figref{fig:drag}), enhancing $f_{\rm B}$ steepens the
metallicity gradient at small $R$ and diminishes it at large $R$.

\begin{figure}
\epsfig{file=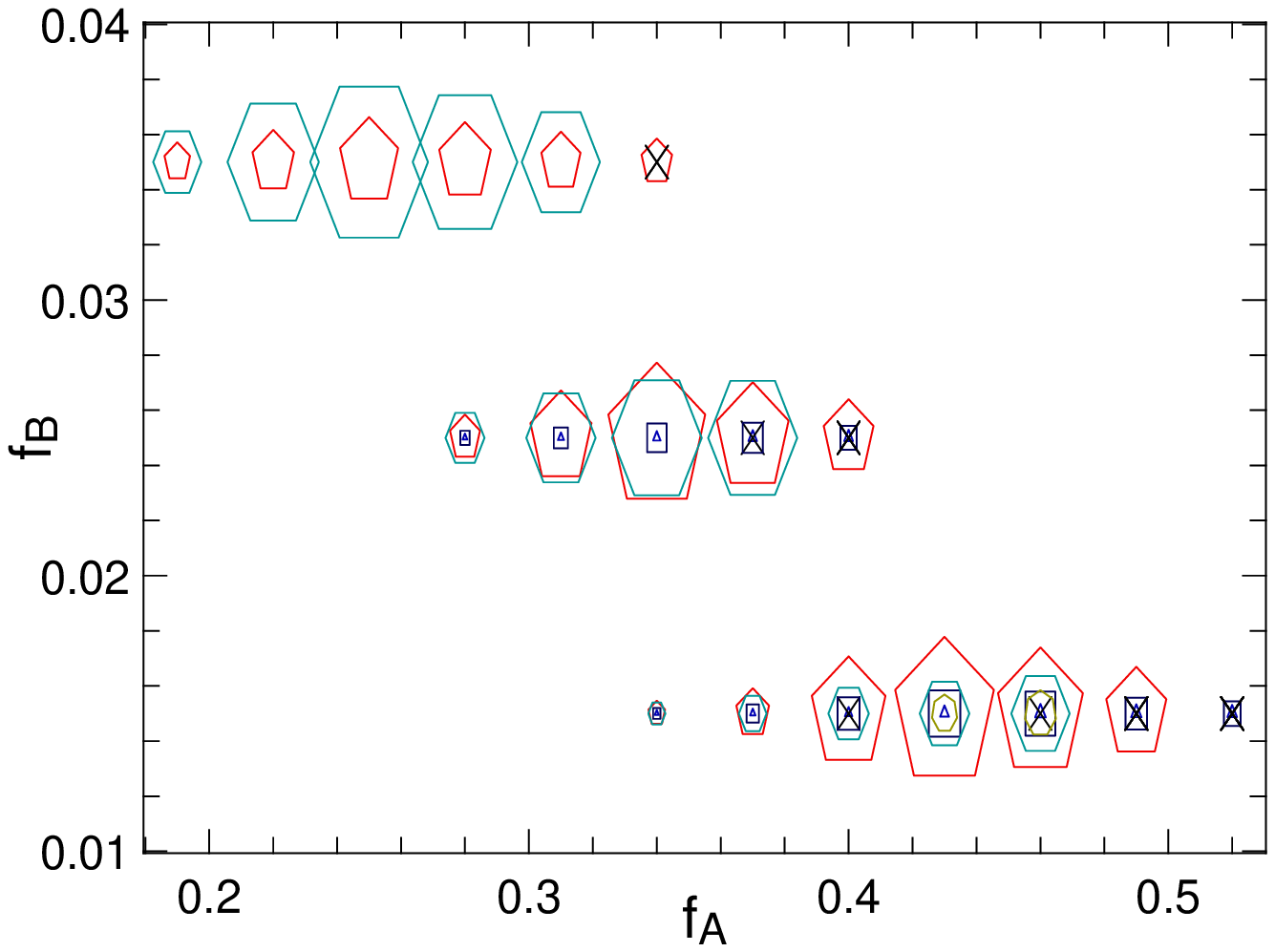,width=\hsize}
\epsfig{file=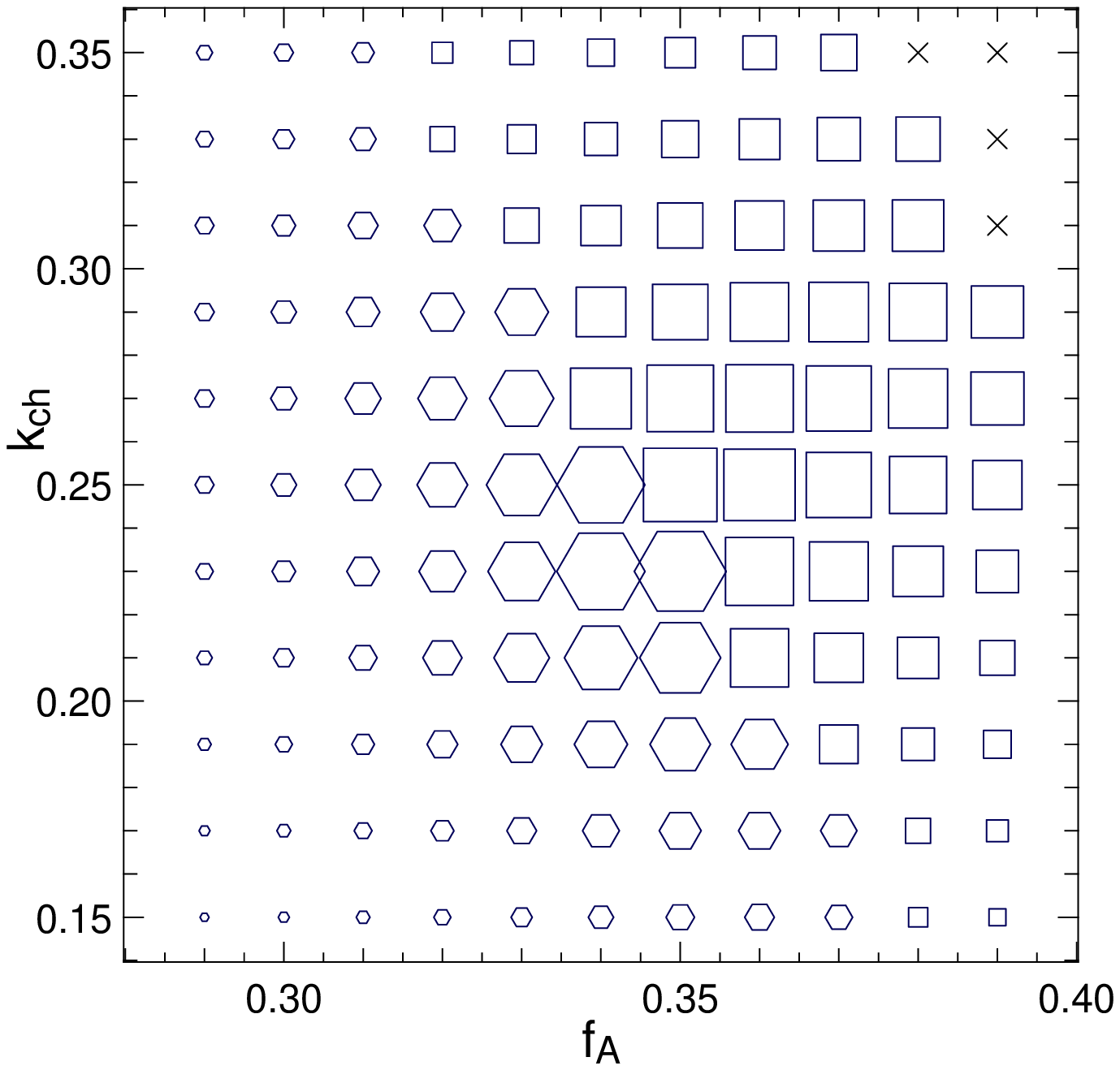,width=\hsize} \caption{The likelihood of the GCS
metallicity distribution in models with infall rates given by equation
(\ref{eq:Mdot}). Upper panel: the values of the infall parameters $f_{\rm A}$
and $f_{\rm B}$ are given by the locations of the symbols, and the value of
$k_{\rm ch}$ is indicated by the number of sides of the polygon: 3, 4, 5,
\dots\ for $k_{\rm ch}=0.0$, $0.1$, $0.2,\ldots$.  The size of the polygon
increases linearly with the log likelihood of the data, models with ages
smaller than $9 Gyr$ are marked with crosses. 
Lower panel: the
likelihoods of models with $f_{\rm B}=0.025$ and varying $(f_{\rm A},k_{\rm
ch}$). In this panel the size of a symbol is a more sensitive function of
likelihood than in the upper panel. Hexagons indicate models with best-fit
ages higher than $12.6\Gyr$, models with best-fit ages below $9.6\Gyr$ are crossed.} \label{fig:aktmods}
\end{figure}

Eliminating churning and radial gas flows (by setting $k_{\rm ch}=0$, $f_{\rm
A}+f_{\rm B}=1$) dramatically reduces the metallicity gradient within
both the stellar and gas discs: at the present epoch the gradient in the gas
near the Sun falls from $-0.11\dex\kpc^{-1}$ to $-0.01\dex\kpc^{-1}$.  Increasing
the churning amplitude $k_{\rm ch}$ both increases the width of the peak in
the predicted solar-neighbourhood metallicity distribution
(Figs~\ref{fig:star_gasZ} and \ref{fig:metdis}) and reduces the gradient in the mean metallicity of
stars at $R\lta R_0$. 

\figref{fig:aktmods} shows how the likelihood of the GCS metallicity
distribution plotted in \figref{fig:metdis} varies with $f_{\rm A}$, $f_{\rm
B}$ and $k_{\rm ch}$. In the upper panel favoured models (with large symbols)
lie along a line that slopes down and to the right. Along this line a
decrease by $0.01$ in $f_{\rm B}$ is compensated by an increase in $f_{\rm
A}$ by $\sim0.08$.

As one moves down the upper panel of \figref{fig:aktmods}, the steepness of
the metallicity gradient near the Sun increases, and the models with $f_{\rm
B}=0.015$ have local gradients steeper than $-0.12\dex\kpc^{-1}$, which may
conflict with the data.  Models higher up the panel have smaller local
metallicity gradients and require larger values of $k_{\rm ch}$ to bring a
sufficient variety of stars to the solar neighbourhood. Models to the right
of the panel have smaller inward flows of gas, leading to local metallicities
that rise faster in time and they match the GCS metallicity distribution at
younger ages, especially if $k_{\rm ch}$ is large so metal-rich stars migrate
to the Sun relatively rapidly.  The structure of the local Hess diagram for
either GCS stars (\figref{fig:GCScomp}) or Hipparcos stars \citep{AumerB08}
implies that the solar neighbourhood is not younger than $9\Gyr$, so when the
age is smaller than $9\Gyr$ the model is marked by a cross in
\figref{fig:aktmods}. Models adjacent to the crosses do not violate the
$9\Gyr$ limit but are nonetheless disfavoured because their local Hess diagrams yield
relatively low likelihoods for the GCS sample.

Four factors make it difficult to confine narrowly the required value of
$k_{\rm ch}$ within the part of \figref{fig:aktmods} that has large symbols:
(i) churning affects mainly the width of the metallicity distribution, which
has less impact on the likelihood than the location of its peak; (ii)
churning, which is strongest in the inner regions of the disc, tends to saturate near the
centre in the sense that old stars become fully shuffled; (iii) the GCS
sample is biased against old and highly dispersed populations of stars, so where churning has the strongest
effect, the observational signature is weak; (iv) the required churning strength
is sensitive to the local metallicity gradient which is not very well
constrained by observations.

In our models there is no azimuthal variation in the metallicity of gas at a
given radius, as is suggested by recent observations \citep[see][]{PrNi08},
which yield very small to negligible inhomogeneity of the ISM at a given
radius.  The effect of relaxing this assumption can be gauged by increasing
the dispersion in the measured metallicities of a given population of stars:
if there is intrinsic dispersion in the metallicity of the ISM in a given
annulus, the measured metallicities of stars formed from it will reflect both
this dispersion and measurement errors. The largest intrinsic dispersion in
the metallicity of the ISM that would appear to be compatible with the data
plotted in \figref{fig:Zgrad} is $\sim0.1\dex$. When we combine this with
measurement errors of $0.1\dex$, we can obtain a fit to the GCS data of
\figref{fig:metdis} that is only slightly worse than that provided by the
standard model by lowering $k_{\rm ch}$ from $0.25$ to $\sim0.1$.

We have studied models with several values of the mass-loss parameter $f_{\rm
eject}$ and concluded that up to the largest values studied ($f_{\rm
eject}=0.15$ at $R<3.5\kpc$ and $0.05$ elsewhere) $f_{\rm eject}$ does not have
a large effect on the model's observable properties, and is anyway degenerate
with the still uncertain nucleosynthetic yields. However, increasing $f_{\rm
eject}$ makes it slightly easier to find an acceptable model, reflecting the
fact that the yields we are using lie at the upper limit of the yields that
are consistent with measured metallicities.

\begin{figure}
\epsfig{file=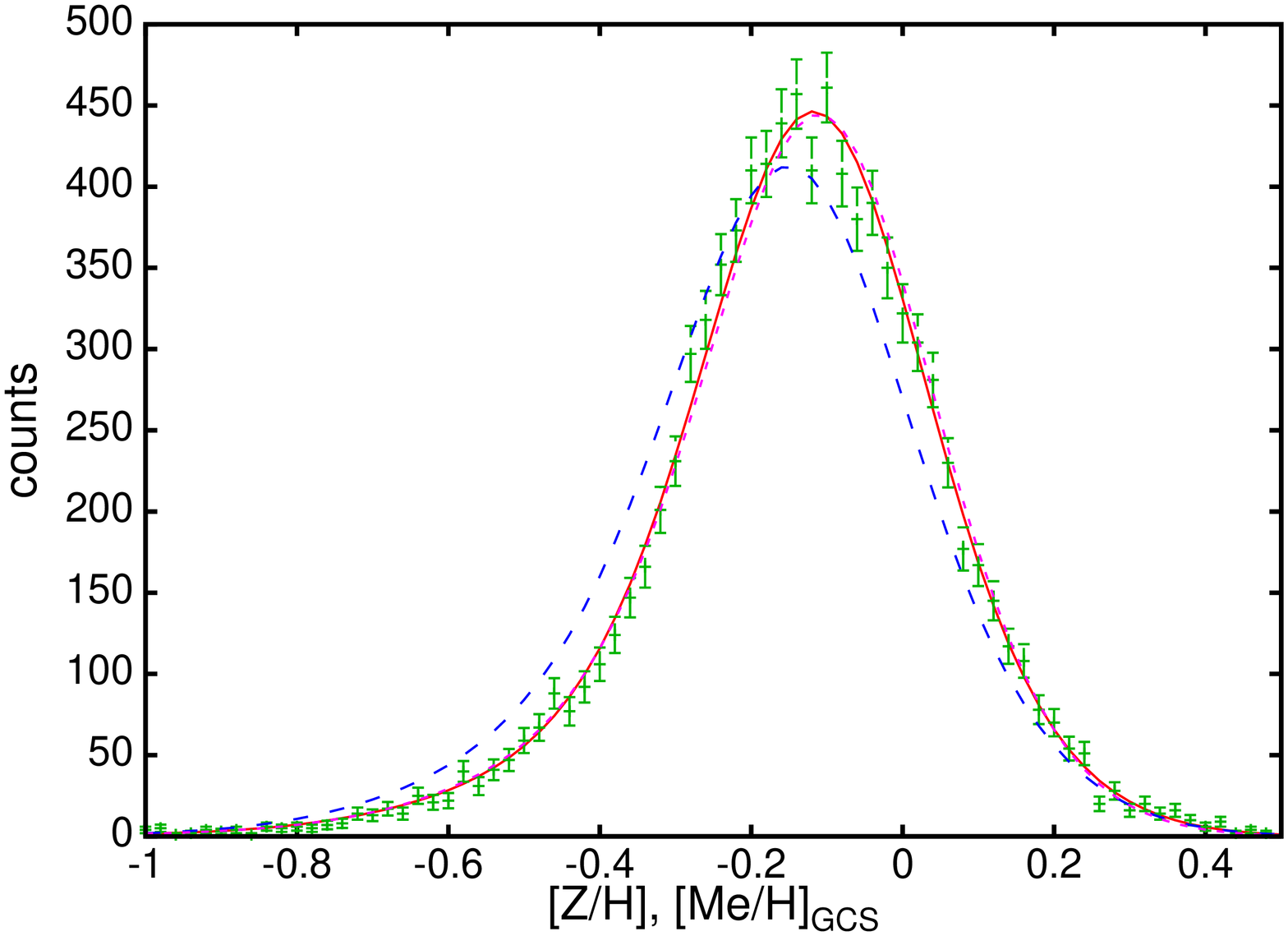,angle=-90,width=\hsize}
\epsfig{file=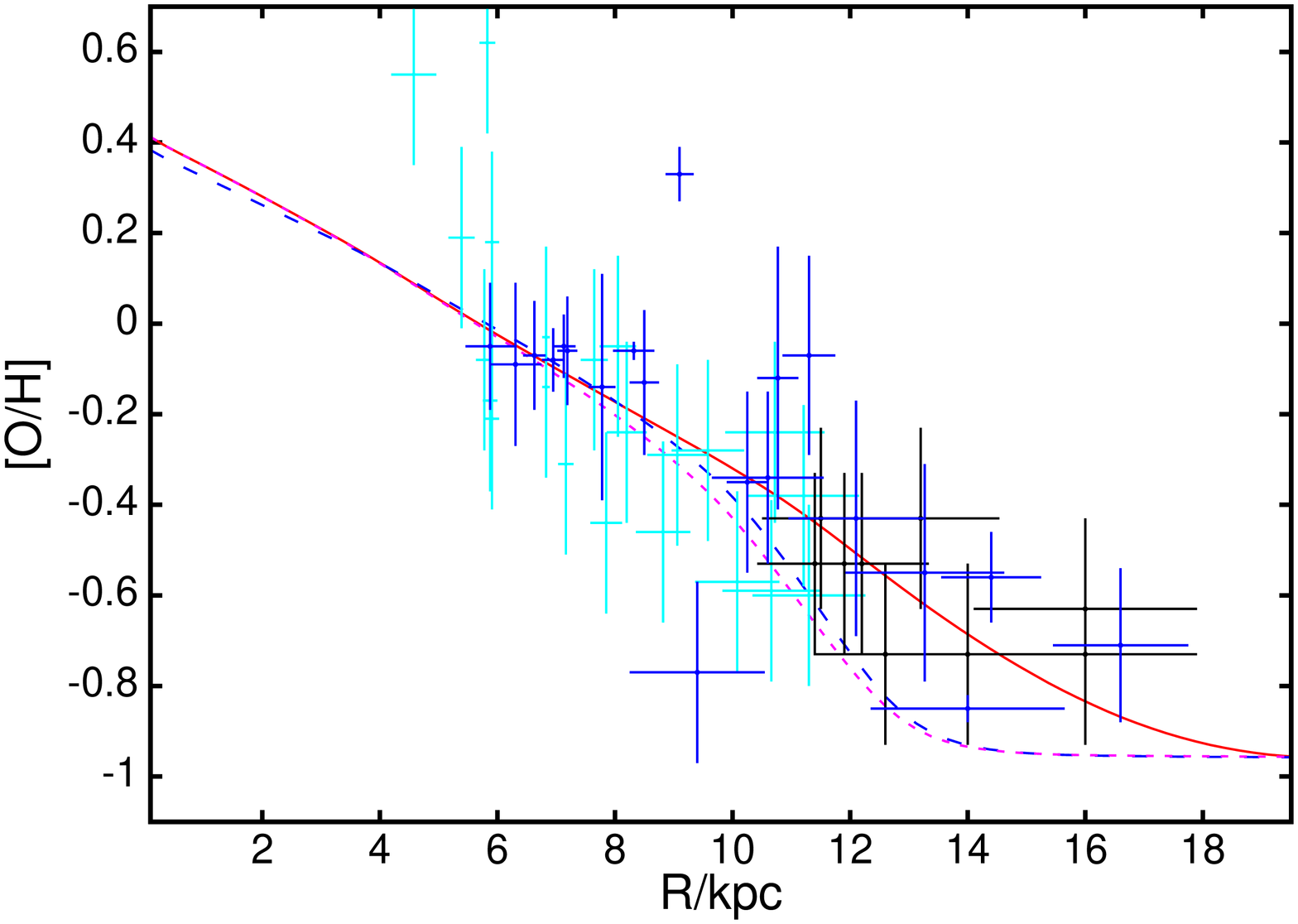,angle=-90,width=\hsize}
\caption{Upper panel: the data points are the GCS counts and the red curve is
the standard model. The broken blue curve shows the effect on this model of raising
$\Sigma_{\rm crit}$ from zero to $2.5\msun\pc^{-2}$. The pink dotted curve
shows that a good fit to the data can be obtained for this value of
$\Sigma_{\rm crit}$. Lower panel: measurements of the metallicity of the ISM
and the predictions of the models shown above.}
\label{fig:Mcut}
\end{figure}

The upper panel of \figref{fig:Mcut} shows the effect on the fit to the GCS
metallicity distribution of using a non-zero value of the threshold gas
density, $\Sigma_{\rm crit}$, below which the SFR declines steeply. Raising
$\Sigma_{\rm crit}$ from zero to $2.5\msun\pc^{-2}$ changes the model
prediction from the red curve of the standard model to the blue curve; the
distribution is now wider and peaks at lower metallicities. The pink dotted
curve shows the result of maximising the likelihood of the data subject to
the constraint $\Sigma_{\rm crit}=2.5\msun\pc^{-2}$. In this model $f_{\rm
A}$ is increased (to $0.44$) and $k_{\rm ch}$ is decreased (to $0.20$)
relative to the standard model. The new model provides a slightly 
worse fit to the GCS data than the standard model, but, as the lower panel reveals,
there is a problem with using a non-zero value of $\Sigma_{\rm crit}$: with
minimal star-formation in the outer disc, the metallicity gradient of the
ISM steepens near the edge of the star-forming regions, while further out the 
metallicity becomes constant at the intergalactic value.
The data show no
sign of this plateau, and are probably incompatible with a plateau as low as
$[Z/\hbox{H}]=-1$. Values of $\Sigma_{\rm crit}>2.5\msun\pc^{-2}$ are
incompatible with the data because they bring the edge of the star-forming
disc too close to the Sun.

By considering the likelihoods of both the Hess diagram and the metallicity
distribution of GCS stars, and our prejudices regarding the proper value of
$f_{\rm eject}$, we chose the model specified by Table~\ref{tab:standard} as
the standard model. With this model the likelihood of the GCS metallicity
peaks at age $11.7\Gyr$.

\section{Relation to other work}\label{sec:reln}

Chemical evolution models of the Galaxy have a long history and a large
literature. It would be inappropriate to attempt to review this literature in
this section. Instead we highlight crucial differences with work that has
most in common with ours, and relate our work to the analysis of the solar
neighbourhood of \cite{Haywood08}.

\subsection{Comparison with earlier models}

Some of the best known models of Galactic chemical evolution are those in
\cite{Chiappini97} and its successors \cite{Chiappini01} and
\cite{Colavitti08}. In each case the disc is made up of annuli that exchange
neither stars nor gas.  Star-formation is driven by a Kennicutt law similar
to equation (\ref{eq:Kenni}). \cite{Chiappini97} introduced a time-dependent
infall rate that is superficially similar to (\ref{eq:Mdot}) but differs from
our infall rate in two important respects. First, the exponential with the
longer time-constant is not turned on until $2\Gyr$ after the start of Galaxy
formation with the consequence that star formation periodically ceases during
the interval $1\Gyr<t<2\Gyr$.  Second, \cite{Chiappini97} make the time
constant $b_2$ a linearly increasing function of radius that vanishes at
$R=0.86\kpc$ \citep[or $1.2\kpc$ in][]{Chiappini01}, whereas here $b_2$ is
constant. If we were to follow the prescription of Chiappini et al., our
inner disc/bulge would become older, and a smaller churning rate would be
required to bring stars more metal-rich than the local ISM to the solar
neighbourhood. An outwards-increasing infall timescale enhances the
metallicity gradient because metallicities are close to their equilibrium
values, and these reflect the ratio of current to past SFRs.

\begin{figure}
\epsfig{file=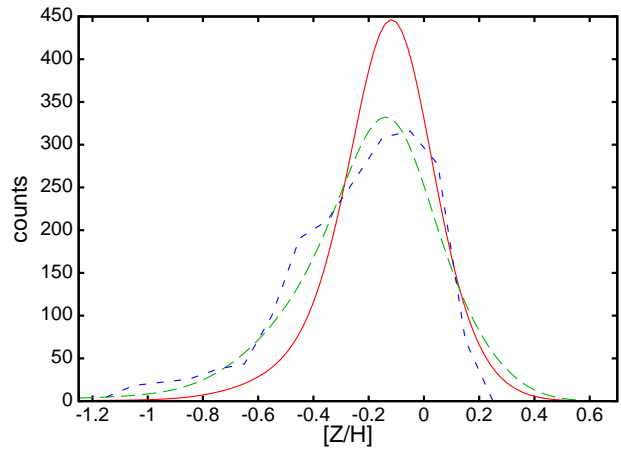,angle=-90,width=\hsize}
\caption{Full red curve: the metallicity distribution predicted by the
standard model for the GCS stars. Long-dashed green curve: the prediction of
the same model for the distribution of the whole solar annulus. Short-dashed
blue curve: the prediction of Chiappini et al.\ (1997) for the solar annulus.}
\label{fig:metdis2}
\end{figure}

\cite{Chiappini97} adjusted their model's free parameters to optimise its fit
to the G-dwarf metallicity distribution for stars in the solar annulus that
was determined by \cite{RochaP96} from 287 stars that lie within $25\pc$ of
the Sun.  \figref{fig:metdis2} illustrates the difference between the
metallicity distributions of stars near the Sun and in the entire solar
annulus: the full red curve shows the standard model's prediction for the
metallicity distribution of GCS stars from \figref{fig:metdis}, while the
long-dashed green curve shows the corresponding distribution in the whole
annulus.  The distribution for the annulus is much broader than that for the
GCS because stars with metallicities far from that of the local ISM are
typically fast-moving and likely to be at high $z$. \cite{RochaP96}
transformed their measured distribution for the local sphere to the modelled
global distribution using correction factors estimated by \cite{SommerL91},
which depend on the local gravitational potential and the velocity
distributions of stars of each metallicity. It is clearly more satisfactory
to use internally generated values of these distributions to predict the
metallicity distribution in the observed volume around the Sun than to infer
the annular distribution from the measured one using external estimates of
the velocity distributions. Moreover, the short-dashed blue curve in
\figref{fig:metdis2} shows the prediction of \cite{Chiappini97} for the solar
annulus. It declines much more steeply at high metallicities than the blue
curve, predicting far too few metal-rich stars. 

The scale of the discrepancy between the blue and the green curves
illustrates that a model that provides an adequate fit to the data of
\cite{RochaP96} is likely to be incompatible with the GCS stars. This
discrepancy is partly due to problems with the calibration of the underlying
dataset: \cite{Haywood02} and \cite{Twarog02} pointed out that the
metallicity calibrations of \cite{Schuster89} underlying those datasets
severely underestimate the metallicities of metal-rich stars. This
underestimation makes the decline in the number of stars at $\feh>0$ steeper
than it should be, and thus makes it easier for the data to be fitted by
traditional models of chemical evolution, which predict a sharp cutoff at
high $\feh$. Apart from its superior calibration, the GCS sample is 50
times larger than that used by \cite{RochaP96}, so its statistical errors are
much smaller and it is a more challenging distribution to fit.

In the models of \cite{Chiappini97}, star formation ceases entirely when the
surface density of gas falls below $\Sigma_{\rm crit}$, while in our models
$\Sigma_{\rm crit}$ merely marks an increase in the rate of decline of the
SFR with decreasing gas density. While a cogent argument can be made for a
rapid decline in the SFR at low gas densities, it is hard to justify a
discontinuity in the rate.  \cite{Chiappini01} conclude that a non-negligible
value of $\Sigma_{\rm crit}$ plays an essential role in fitting the data. In
fact, they set $\Sigma_{\rm crit}=7\msun\pc^{-2}$, with the consequence that
star formation in the solar neighbourhood was constantly stopping and
starting, both during the first $2\Gyr$ and the last $4\Gyr$ of Galactic
history (see their Fig.~4). Since the metallicity of the ISM declines when
star formation has been switched off, this erratic behaviour broadens the
stellar metallicity distribution.  In our models the SFR is steady and in
fact setting $\Sigma_{\rm crit}=7\msun\pc^{-2}$ would result in almost no
stars forming at $R\gta10\kpc$ because the flow of gas through the disc would
prevent the surface density building up to $7\msun\pc^{-2}$.  Hence in our
models  star-formation can occur at large radii only if $\Sigma_{\rm crit}$ is
set to a small or vanishing value. An important difference between our models
and those of \cite{Chiappini01} is that in our models the SFR is a smooth
function of time and the bimodality in $[\alpha/\hbox{Fe}]$ is achieved
without a dip in the star-formation rate.

 \cite{Chiappini97}  give for the inner and outer parts of their models the
radial gradients in the abundances of several elements. These gradients are
largest in the inner regions, but even there they are much smaller than in
our models: for example at $12\Gyr$ the inner gradients of [O/H] and [Fe/H] are
$-0.023$ and $-0.027\dex\kpc^{-1}$ compared to values $\sim-0.08\dex\kpc^{-1}$
and $\sim-0.11\dex\kpc^{-1}$
obtained here. Our larger gradients are a direct consequence of the advection
inwards of the products of nucleosynthesis. The GCS stars show a similar
gradient in $\zeh$: \cite{HolmbergNA} derive a gradient of
$-0.09\dex\kpc^{-1}$.

\cite{Colavitti08} used infall rates measured from simulations of clustering
cold dark matter (CDM) in slightly updated models of
\cite{Chiappini97}. No empirically determined infall rate gave such
satisfactory results as the earlier double-exponential rate. The models were
again compared to the metallicity distribution of \cite{RochaP96} and in most
cases provided inadequate fits to the data. The empirically measured infall
rates are very irregular in time, with the result that the model experiences
powerful bursts of star formation. These lead to large excursions in the
predicted plots of [O/Fe] versus [Fe/H] for which there is no evidence in the
data. Moreover, studies of the past SFR in the disc show no signs of major
bursts of star formation a few gigayears ago.
The disappointing results of this study suggest that the rate at which
gas joins the disc is not simply the ratio of masses of baryons and
dark-matter times the dark-matter accretion rate. In fact, the wealth of
evidence that the majority of baryons are still in the intergalactic medium
\citep{Persic,Fukugita98} is a clear indication that galaxies do not acquire gas as
fast as this naive calculation suggests; rather gas is stored in the warm-hot
intergalactic medium (WHIM) and from there accreted at a still uncertain
rate. It seems unlikely that chemical evolution models will be successfully
coupled to cosmological clustering simulations until we have understood the
complex interface between the WHIM, cold infall and galactic fountains.

\cite{NaabO06} determined the infall rate by assuming that the surface
density of the disc is always exponential, but with a scale length that is
proportional to $v_{\rm c}/H$, where $v_{\rm c}$ is the Galaxy's circular
speed (taken to be constant) and $H(t)$ is the Hubble parameter. The infall
rate at all times and places was fixed by assuming that the central surface
density of the disc is constant at its current value. The model's observables
were calculated by using the Kennicutt law (\ref{eq:Kenni}) to convert gas to
stars. A very simple approach to chemical evolution was employed, in which
material does not move between annuli and only the overall metal content $Z$
was followed. In these models the metallicity gradient in the ISM becomes
less steep over time, and is now $\sim-0.04\dex\kpc^{-1}$. Their model
predicts for the solar neighbourhood fewer metal-rich stars and  more
metal-poor stars than are in the GCS.

In all these models, solar-neighbourhood stars should satisfy a well defined
metallicity-age relation, and the G-dwarf metallicity distribution is simply
the result of combining this relation with the SFR-age relation. As we have
seen, the observed width of the local metallicity distribution is
approximated by exploiting irregular time evolution of the metallicity of the
local ISM \citep[cf.\ also][]{Portinari98}. As Fig. ~\ref{fig:alpha_thick}
illustrates, our models solve this problem in an entirely different way, and
using a simpler history of star formation.

\subsection{Haywood's analysis of the solar neighbourhood}\label{sec:Haywood}

\cite{Haywood06} and especially \cite{Haywood08} has critically re-examined
the age-metallicity distribution of the GCS stars and concluded that the data
are only consistent with the existence of a well defined age-metallicity
relation for disc stars younger than $\sim3\Gyr$; for such young stars the
spread in [Fe/H] is consistent with the expected dispersion in the
metallicity of interstellar gas at a given radius. At ages larger than
$3\Gyr$ the width of the metallicity distribution is larger than can be
accounted for by measurement errors and inhomogeneity of the ISM. In
particular, there are stars with ages $\sim5\Gyr$ that have
$\hbox{[Fe/H]}=0.5$, and stars with ages $<7\Gyr$ that have
$\hbox{[Fe/H]}=-0.5$. The older the age bracket that one examines, the wider
the range of metallicities present. Our predicted age-metallicity
distribution (\figref{fig:ISMZ}) is in good agreement with that derived by
\cite{Haywood08}.

\cite{Haywood06} finds that when his revised ages are used for GCS stars, the
metallicities of thick disc stars increase as their ages decrease. That is,
these stars point to rapid self-enrichment of the thick disc. In fact,
Haywood argues that the thick disc is not the relic of some captured
satellite(s) but an integral part of the Galaxy's disc and has played a
central role in the chemical evolution of the thin disc. Chemical evolution
models should treat the disc as a whole, not just individual parts.  Our
results strongly underline this conclusion from a theoretical perspective.
The metallicity-age plot shown in Fig.~1(b) of \cite{Haywood08} is
satisfyingly similar to the lower panel of our \figref{fig:ISMZ}: in both
figures the stellar density is highest around ($\tau=2\Gyr,\feh=-0.1$) and in
this region the ridge line gradually drops to the right. At older ages the
distributions become broader, being confined by $\feh\sim0.4$ and $-0.5$. At
ages greater than $10\Gyr$ both distributions reach down to $\feh=-1$. In
fact the small differences between our figure and that of Haywood are readily
accounted for by the substantial errors in measured stellar ages. This fit is
remarkable because the model parameters were chosen without reference to
measured stellar ages.

An argument sometimes advanced for a dichotomy between the thick and thin
discs is the existence of two sequences in the
$([\alpha/\hbox{Fe}],\hbox{[Fe/H]})$ plane (\figref{fig:alpha_thick}). This
diagram suggests that the last thick-disc stars to form had higher abundances
than the first thin-disc stars to form. To explain this finding in the
context of a conventional chemical-evolution model, sudden dilution of the
ISM by a massive gas-rich accretion is required \citep{Bensby05,Reddy06}.
Against this proposal \cite{Haywood06} objects that the bulk of the oldest
thin-disc stars have $\hbox{[Fe/H]}\simeq-0.2$ and there is no evidence that
the most metal-poor thin-disc stars are particularly old. Our models show
that the observed structure of the $([\alpha/\hbox{Fe}],\hbox{[Fe/H]})$ plane
arises naturally when radial migration is allowed. \cite{Haywood08} examined
the orbital parameters of stars of various metallicities and showed that
local thin-disc stars with metallicities that overlap the metallicity range
of the thick disc have higher angular momenta than more typical thin-disc
stars. Similarly, he found that stars in the high-metallicity tail of the
local metallicity distribution have low angular momenta. Even though churning
could in principle eradicate the correlation between angular momentum and
metallicity, our models reproduce these correlations
(Figs~\ref{fig:alpha_thick} and \ref{fig:vphiZ}) because blurring makes a
sufficiently large contribution to bringing these chemically anomalous stars
near the Sun.

\cite{Ivezic08} also argue that in the SDSS data the kinematic properties of
the thick disc evolve continuously with distance from the plane in a way that
suggests that the thick disc joins continuously to the thin disc. By
contrast, \cite{Veltz} argued for a clean break between the thin and thick
discs on the basis of shallow local minima in the density of 2MASS stars
$n(z)$ in seen at the Galactic poles as a function of photometric distance
$z$.  As \cite{Veltz} show, minima in $n(z)$ are not expected if the disc is
a superposition of two exponential structures, but the minima yield a clean
discontinuity in the distribution of velocity dispersions when
multi-component isothermal distribution functions are used to model the data.
The very unexpectedness of the minima makes the modelled break extremely
clean. It will be interesting to see whether the distribution of measured
radial velocities of stars in the RAVE survey substantiate these model
velocity distributions.

\section{Conclusions}\label{sec:conclude}

It is now more than forty years since the theory of stellar evolution
attained the level at which it became possible to model the chemical
enrichment of the ISM. From the beginning of that endeavour measurements of
the abundances of individual solar-neighbourhood stars have played a key role
because a star preserves like a time capsule the state of the ISM at
the remote epoch of its formation. Considerable theoretical and observational
efforts have been devoted to probing the history
of the Galaxy with this connection.

Half a century ago, \cite{Roman50,Roman54} and others discovered the
connection between the kinematics and chemistry of stars, yet curiously
little has been done to include kinematics in models of chemical evolution.
The general presumption has been that each annulus of the disc evolves
independently of others, and the well-known correlations between chemistry
and kinematics can be understood as arising through the stochastic
acceleration of stars: older stars tend to have larger random velocities and
lower metallicities. No effort was made to develop greater diagnostic power
by simultaneously modelling chemistry and kinematics.

The continued use of mutually independent annuli by modellers of chemical
evolution is surprising given that it was from the outset recognised that
many stars are on significantly non-circular orbits that each radial
half-period cover more than a kiloparsec in radius. In fact, it has generally
been assumed that the radial velocity dispersion within the disc rises as one
moves inward to values that lead to radial excursions of several kiloparsecs
(see \figref{fig:blurr}). Moreover, observations have long indicated that
galactic discs have significant metallicity gradients
\citep[e.g.][]{Vila-CostasE92,deJong96}, so radial migration of stars is
bound to leave a signature on the metallicity distribution of
solar-neighbourhood stars. We have called this aspect of radial migration
``blurring''.

The present study owes its impetus to the discovery by \cite{SellwoodB} that
radial migration is a more potent process than mere blurring: the dominant
effect of transient spiral arms is not to heat the stellar disc as had been
supposed, but to cause stars either side of corotation to change places
without moving to eccentric orbits (``churning''). \cite{SellwoodB} did not
demonstrate that gas participates in churning, but they argued that it must
on dynamical grounds, and \cite{Roskar1} found evidence that this was the case
in their N-body--SPH simulations of galaxy formation.  Since churning is an
aspect of spiral structure that can {\it only\/} be probed through its impact on
chemical evolution, we wanted chemical-evolution models that included
churning, and logically it was natural to extend these models to include both
blurring and radial gas flows. 

An ingredient of our models that might be controversial is the introduction
of radial gas flows. We have absolutely no reason to expect that the infall
profile is exponential, so if discs are to be exponential this must be the
result of flows through the discs redistributing mass within the disc. In
fact, as gas streams through spiral arms it dissipates energy in shocks that
is ultimately gravitational energy that becomes free as the gas surrenders
angular momentum to the stars and drifts inwards. Hence at some level inward
gas flows are mandatory \citep{LaceyF85}. 

Unfortunately, the theory of galaxy formation has yet to advance to the point
at which it can prescribe the spatial and temporal structure of gas
accretion, so it is necessary to parametrise accretion in some way. The
accretion process must be constrained to result in the formation of the
observed stellar and gaseous discs. Our accretion Scheme AB satisfies this
constraint for all values of the parameters, but it is inevitably acausal
in that the formation mechanism is being driven by its known outcome. While
its acausality is unattractive, Scheme AB is a flexible parametrisation that
enables us to form exponential discs for a variety of different assumptions
about the radial density of infalling gas, and the resulting radial profiles
of infall and gas-flow (\figref{fig:drag}) are entirely plausible.

The impact that radial migration has on the local metallicity distribution
obviously varies with the magnitude of the metallicity gradient in the ISM,
which in turn depends on the gas flow within the disc and therefore the
radial infall profile. For this reason the most important parameters of our
models are $f_{\rm A}$ and $f_{\rm B}$, which control the distribution of
infalling gas.

The models provide good fits to the GCS counts of stars as functions of
$\feh$, $M_V$, $T_{\rm eff}$ and stellar age, as well as reproducing the
correlations between tangential velocity and abundance patterns that have
been pointed out by \cite{Haywood08}. These fits are achieved by churning,
blurring and radial flows working together. They depend on the existence of
an appreciable metallicity gradient in the ISM, which is established by the
radial flow of gas, and they depend on radial mixing of stars by blurring and
churning. The steeper the metallicity gradient in the gas, the smaller the
effect of churning can be, but for any observationally consistent metallicity
gradient, churning has a non-negligible role to play.

The models describe the coevolution of the thick and thin discs, and presume
that thick-disc stars were formed in the Galaxy rather than accreted from
outside. Given the simplicity of our assumptions, the extent to which a
dichotomy between an $\alpha$-enhanced thick disc and a solar-type thin disc
automatically manifests itself in the models is remarkable. In particular, in
the solar annulus the distributions of [O/H] at given $\feh$ are bimodal in
the range of $\feh$ associated with the overlap of the two discs
(\figref{fig:alpha_thick}), the vertical density profile can be represented
at the sum of two exponentials, and at $R<R_0$ the radial density profile
becomes flatter at lager distances from the plane (\figref{fig:thick_thin}).
None of these characteristics is dependent on our choice of a double
exponential for the time dependence of infall: a model in which the gaseous
mass, and therefore star-formation rate, is held constant also has these
features. They are consequences of the $\sim1\Gyr$ timescale of type Ia
supernovae and the secular heating and churning of the disc.

The models assume that scattering of stars increases the velocity dispersion
of a coeval population as $t^{1/3}$, but the models go on to predict that 
within the solar neighbourhood  velocity dispersion increases as a higher
power of age, roughly $t^{0.45}$ in agreement with what is found from
Hipparcos stars with good parallaxes. This finding may reconcile scattering
theory, which cannot readily explain an exponent in excess of $1/3$, with
observations. The key point is that radial mixing brings to the solar
neighbourhood stars born at small radii, where the velocity dispersion is
undoubtedly large.

The nucleosynthetic yields from each generation of stars are still
significantly uncertain. Our philosophy has been to use standard values from
the literature rather than exploit uncertainties in the yields to tune the models
to the data. The yields we are using are in the upper region of those that can
provide adequate fits to the data, with the consequence that increasing the
value of the mass-loss parameter $f_{\rm eject}$, which is degenerate with
the magnitude of the yields, makes it easier to fit the data.  Our yields are
surely not exactly right and consequently some of the properties the models
derive from them will be in error. On account of these uncertainties we have
suppressed the predictions of our code for the abundances of certain
elements, most notably carbon.

The discovery that three-dimensional, non-equilibrium models of the solar
atmosphere require the metal abundance of the Sun to be $Z_\odot=0.012-0.014$
\citep{Grevesse07} poses a major problem for this field. A prerequisite for
successful chemical modelling is a consistent metallicity scale for both
stars and the ISM. At present the only consistent scale is the traditional
one on which $Z_\odot=0.019$, so this is the one we have used.

If a new scale were established on which all metallicities were significantly
lower, viable models could be produced by lowering the yields. A
straightforward way to do this would be to lower the maximum mass in the IMF:
reducing this mass from $100\msun$ to $50\msun$ would reduce yields by
$\sim30$ percent in line with the proposed reduction in $Z_\odot$. The oxygen
yield would come down fastest, reducing [O/Fe] by $\sim0.2\dex$.

Although we believe that this study represents a significant advance on all
previous models of Galactic chemical evolution, it is highly imperfect. Some
major weaknesses of our work are the following.

\begin{itemize}
\item[1.] We have  assumed that the probability of mass interchange between
rings is proportional to the product of the rings' masses. This probability
reflects the number and intensity of spiral feature with corotation at that
radius, and should be a function of both mass and velocity dispersion. A
further study of self-gravitating discs similar to that of \cite{SellwoodB}
would be necessary to determine this function.

\item[2.] We have assumed that the vertical and radial motions of stars
decouple. This assumption has a significant impact on both the relation
between age and velocity dispersion in the solar neighbourhood and on the
predicted vertical density profile in the solar annulus, which is interesting
in itself and impacts on the selection function of GCS stars and thus on our
choice of standard model.  The assumption is unjustifiable for stars on
eccentric orbits, which do play an important role in the model fits.
Unfortunately, a sounder treatment is impossible until a better approximation
to the third integral of Galactic dynamics is available. It is our intention
to resolve this problem through ``torus modelling''
\citep[e.g.][]{McmillanB08}.

\item[3.] Our models include radial mixing of gas through churning and
viscous inspiralling, but do not include radial redistribution of gas by the
Galactic fountain \citep[e.g.][]{Benjamin97}. A significant body of evidence
indicates that star formation drives neutral hydrogen to kiloparsec heights
above the plane. NGC$\,$891, which is not dissimilar to the Galaxy, has
$\sim25$ per cent of its neutral hydrogen more than $1\kpc$ from the plane
\citep{OosterlooFS}.  This extraplanar gas must move over the plane on nearly
ballistic trajectories, and in the absence of interaction with the corona
(gas at the Galaxy's virial temperature $\sim2\times10^6\K$), the gas must
return to the plane further out than its point of ejection. However,
observations suggest that neutral gas above the plane is actually flowing
inward rather than outward, presumable as a result of interaction with the
corona \citep{FraternaliB08}. The mass of extraplanar gas is so large and the
timescale for its return to the plane so short that whichever way this gas
flows, it has a considerable potential for radially redistributing metals.
Extraplanar gas must be ejected from the disc by supernova-heated gas that is
probably highly metal-rich. Some of this gas will be lost to the Galaxy as we
have assumed, but some of it will return to the plane with infalling gas.
Again there is potential here for significant radial redistribution of metals
that has been ignored in the present study.

\item[4.] Our treatment of the inner Galaxy is unacceptably crude in that we
have replaced the bar/bulge with a disc. Unfortunately introducing the bar
opens a Pandora's box of complexities, and at the present time is probably
only feasible in the context of a particle-based model such as those of
\cite{SamlandG} and \cite{Roskar2}. Our hope is that our imaginary central
disc has a similar impact on the chemodynamical state of the local disc to
the combined impact of the giant star-forming ring at $R\simeq4\kpc$, the
gas-deficient region interior to this ring and the star-forming $x_2$ disc at
$R\lta200\pc$.

\end{itemize}

From this list of shortcomings of our models it is clear that we are still
far from a definitive account of the Galaxy's chemical evolution. We shall be
satisfied if we have convinced the reader that the interplay between dynamics
and chemistry is so tight at to be indissoluble. This fact is at one level
inconvenient because it undermines the value of conclusions drawn from
traditional models of chemical evolution. But at another level it represents
an opportunity to learn more. The connection between the kinematics of stars
and the compositions of stars and gas, which can be measured in great detail,
involves three areas about which we are too ignorant: the distribution of dark
matter within and around the Galaxy, the Galaxy's history of assembly, and
the nature of the Local Group's  IGM. By building dynamical models of  the
Galaxy that have chemical evolution built in to the basic structure, we
should be able to make decisive progress with one of the major problems of
contemporary astronomy.

We thank M. Haywood, J.-U. Ness and A. Riffeser for fruitful discussions.
R.S. acknowledges material and financial support from the Studienstiftung des
Deutschen Volkes and Stiftung Maximilianeum and the hospitality of Merton
College Oxford, where this work began.

\begin{figure}
\epsfig{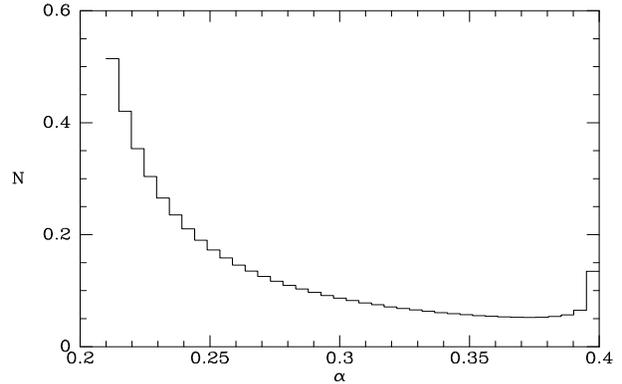}
\caption{The distribution of $\alpha$ abundances predicted by equation
(\ref{eq:Malpha}).\label{fig:App}}
\end{figure}
\section*{Appendix: origin of bimodal [O/Fe] distributions}

We provide an analytic model of the development of the bimodal distributions
of [O/Fe] evident in Fig.~\ref{fig:alpha_thick}. We assume that star
formation starts at $t=0$ and that the SFR is $\propto\e^{-Kt}$ for $t>0$.
Consistent with equation (\ref{eq:SN}), we assume that a coeval group of
stars formed at $t'$ generates a rate of type Ia supernovae that vanishes for
$t<t'+t_0$ and is subsequently $\propto\e^{-k(t-t')}$. Then given that the rate of
core-collapse SNe is proportional to the SFR, the rate of production of Fe
is
 \begin{equation}
{\d M_{\rm Fe}\over\d t}=\cases{b\e^{-Kt}&for $t<t_0$\cr
b\e^{-Kt}+c\int_0^{t-t_0}\d t'\,\e^{-Kt'}\e^{-k(t-t')}&for $t\ge t_0$,}
\end{equation}
 where $b$ and $c$ are constants. Integrating this production rate, we obtain the
 iron mass at time $t>t_0$ as
\begin{eqnarray}\label{eq:Malpha}
M_{\rm Fe}(t)&=&b{1-\e^{-Kt'}\over K}\\
&&+{c\over
k-K}\left(\e^{-(k-K)t_0}{\e^{-Kt_0}-\e^{-Kt}\over K}-{\e^{-kt_0}-\e^{-kt}\over
k}\right).\nonumber
\end{eqnarray}
 In the
approximation that SNIa do not contribute $\alpha$ elements, and that the
delay in the production of these elements by a stellar population is
$\ll1/k$, the mass of $\alpha$ elements is $M_\alpha=(1-\e^{-Kt})a/K$ where $a$ is a
constant. Then equation (\ref{eq:Malpha}) predicts that $\alpha\equiv
M_\alpha/M_{\rm Fe}$ equals $a/b$ for $t\le t_0$ and then drops rapidly
towards its asymptotic value,
\begin{equation}
\alpha(\infty)={b\over K}+{c\over Kk}\e^{-kt_0}.
\end{equation}
 Fig.~\ref{fig:App} plots the distribution at $t=13\Gyr$ of stars over
$\alpha$ when the initial and asymptotic values of $\alpha $  are set to
$0.4$ and $0.2$ and the other parameters are $K=1/7\Gyr$, $k=1/1.5\Gyr$ and
$t_0=0.3\Gyr$, which allows $0.15\Gyr$ for white dwarfs to form and
$0.15\Gyr$ for them to accrete prior to deflagrating.

\label{lastpage}

\end{document}